%% file: arxiv.tex
\documentclass[sigconf,screen,authorversion,nonacm,balance=false]{acmart}

\usepackage{fancyhdr}
\AtBeginDocument{%
    \addtolength{\footskip}{2.0\baselineskip}
    \fancyfoot[L]{\textbf{\textit{© Philipp Danner 2024. This is the author's version of the work. It is posted here for your personal use. Not for redistribution. The definitive version was published in Proceedings of the Fifteenth ACM International Conference on Future Energy Systems (e-Energy '24), \url{https://doi.org/10.1145/3632775.3661986}.}}}
}

\usepackage{xcolor}
\definecolor{cb_1}{rgb}{0.902, 0.380, 0.004}
\definecolor{cb_2}{rgb}{0.992, 0.722, 0.388}
\definecolor{cb_3}{rgb}{0.698, 0.671, 0.824}
\definecolor{cb_4}{rgb}{0.369, 0.235, 0.600}

\usepackage{hyperref}
\usepackage[nolist]{acronym}
\usepackage{tabularx}
\usepackage{multirow}
\usepackage{subcaption}

\usepackage[ruled,vlined,commentsnumbered]{algorithm2e}

\SetCommentSty{mycommfont}

\begin{document}

\title{Two-Step Blackout Mitigation by Flexibility-Enabled Microgrid Islanding}

\author{Philipp Danner}
\email{philipp.danner@uni-passau.de}
\orcid{0000-0002-3005-630X}
\affiliation{
  \institution{University of Passau}
  \city{Passau}
  \state{Bavaria}
  \country{Germany}
}
\author{Anna Volkova}
\email{anna.volkova@uni-passau.de}
\orcid{0000-0001-9198-265X}
\affiliation{
  \institution{University of Passau}
  \city{Passau}
  \state{Bavaria}
  \country{Germany}
}
\author{Hermann de Meer}
\email{demeer@uni-passau.de}
\orcid{0000-0002-3466-8135}
\affiliation{
  \institution{University of Passau}
  \city{Passau}
  \state{Bavaria}
  \country{Germany}
}

\input{sources/00_abstract.tex}
\keywords{microgrid islanding, consensus algorithm}
\maketitle{}

\input{sources/acronyms}
\input{sources/01_motivation.tex}
\input{sources/02_methodology.tex}
\input{sources/03_evaluation.tex}
\input{sources/04_conclusions.tex}

\begin{acks}
  This project has received funding from the European Union's Horizon 2020 research and innovation programme under grant agreement No 957819.
  It has also received funding from Deutsche Forsch\-ungsgemeinschaft (DFG, German Research Foundation), project number 360475113, as part of the priority program DFG SPP 1984 - Hybrid and Multimodal Energy Systems: System theory methods for the transformation and operation of complex networks.
\end{acks}

\bibliographystyle{ACM-Reference-Format}
\bibliography{refs.bib}

\input{sources/appendix.tex}

\end{document}

%% file: sources/00_abstract.tex
\begin{abstract}
     Blackouts are disastrous events with a low probability of occurrence but a high impact on the system and its users. With the help of more distributed and controllable generation and sector-coupled flexibility, microgrids could be prepared to operate in islanded mode during a blackout.
     This paper discusses a two-step blackout mitigation approach for highly renewable microgrids that utilizes user flexibility and energy storage systems for power balance in islanded grid operation. The proposed method includes a proactive flexibility reservation step, which derives a minimal reservation schedule for microgrid resources under uncertainty considering related operational costs. As a second step, during a blackout, a fully distributed control is implemented to maximize the usage of available resources based on a sequence of max and min-consensus rounds.
     This paper focuses on the second step, for which the effectiveness of blackstart and long-term coordination is shown. Load shedding can be reduced by 40\% compared to the forecast value. A hardware-in-the-loop simulation of a grid-forming converter further showed a fast convergence toward the optimal operation point.
\end{abstract}

%% file: sources/acronyms.tex
\begin{acronym}
    \acro{MG}{Microgrid}
    \acro{SoC}{State of Charge}
    \acro{MPC}{Model Predictive Control}
    \acro{MAS}{Multi Agent System}
    \acro{GFR}{Grid-Forming Resource}
    \acro{ESS}{Energy Storage System}
\end{acronym}

%% file: sources/01_motivation.tex
\section{Motivation}
Triggered by the ongoing energy transition from fossil to renewable generation, two significant trends can be observed: (1) decentralization introduced by distributed small-scale generation and flexibility units, such as photovoltaic, wind, or biomass power plants, and battery storage systems, and (2) digitization and thus controllability of an increasing number of those distributed generation and flexibility units.
These trends enable new organizational concepts to manage the energy system within regional proximity.
The technical concept focused on power systems is called a \ac{MG}, a self-contained cluster of resources connected by a medium or low voltage network~\cite{smith2013}. \acp{MG} can optimize collective self-consumption or aggregate flexibility for energy or ancillary service markets~\cite{ManuelMauricio2021}.
The massive deployment of converter-interfaced renewable energy sources, on the other hand, increases the variability of generation forecasts and lowers the system inertia, which potentially could lead to wide-area blackouts due to critical load situations~\cite{HaesAlhelou2019}. The connectivity of distributed generation further increases the attack surface for cyber-attacks that may cause blackouts~\cite{Fotis2022}.
In such low-probability disaster situations, which highly impact the system and its users, \acp{MG} can operate in islanded mode~\cite{Wang2016a}.

The islanded operation mode poses hard constraints to the system as consumption and generation must be balanced on all time scales. In a highly renewable future energy system, grid-forming services can be provided by \acp{GFR}, such as power electronic-based converters connected to a renewable energy source or storage system~\cite{Lasseter2020}, instead of conventional fossil-driven spinning generators. Distributed flexibility plays a significant role in long-term power balance, which involves more energy. The core of this paper is the coordination of this flexibility to enable emergency power supply for high-priority loads, such as the communication system, fire brigade, or a medical facility. In parallel, additional loads should be dynamically powered in times of renewable energy oversupply.

Energy management in \acp{MG} is classified into (1) \textit{planning} (years, months ahead) and (2) \textit{operation} (day, hours ahead up to hours to days after a certain event). Operation can be further distinguished by the time of the event occurrence into (2a) \textit{\ac{MG} scheduling} or resilience-oriented response (day, hours ahead), and (2b) \textit{\ac{MG} islanded control} or power restoration (minutes up to days after the event)~\cite{Mahzarnia2020,Anderson2020}, which relate to the two steps in this research work. Our main focus is on islanded control, which is discussed in more detail. \ac{MG} planning, including generation and storage sizing and placement, and system-wide power restoration are not covered.

\subsection{Related Work}
For \ac{MG} islanded control, frequency-watt droop-based methods for instantaneous to short-term control are proposed in~\cite{Mahmood2022, Lu2022} and compared in~\cite{Vandoorn2013}.
Droop-based control, however, cannot handle discrete flexibility with priorities such as detachable loads while guaranteeing constant frequency~\cite{Vandoorn2013}. Furthermore, control over frequency is proposed for balancing power between multiple \acp{GFR}~\cite{Sadeque2021}.

Communication-based methods with bidirectional information exchange enable a more granular and potentially more optimal control on longer timescales~\cite{espina2020controlstrategies}. Distributed algorithms are commonly applied in a \ac{MAS} as they are not dependent on the availability of a central control unit and have high adaptability to the unpredictable \ac{MG} structure and asset availability during a blackout~\cite{Rokrok2017, stark2021restoration, zidan6205352mas}. Accurate distributed coordination is required to decide on the switching sequence in a \ac{MG}. In this work, a consensus-based control strategy is discussed.

Consensus protocols can be categorized as synchronous or asynchronous, and the literature discusses different aspects of them, such as convergence~\cite{Giannini2016, Lin2007, Shi2015}, max-plus algebra~\cite{Nejad2009}, or signal noise~\cite{Zhang2016}. Consensus problems can also be classified as static or dynamic. Static problems involve a constant signal, while dynamic problems involve a varying signal and/or number of agents. Although there are methods to approximate the maximum of dynamic signals~\cite{Deplano2023, Monteiro2020}, this paper models varying input signals as a sequence of static consensus problems.
Consensus-based control has already been demonstrated in the context of voltage regulation~\cite{Danner2022}, energy utilization efficiency~\cite{Zhou2023}, controlled islanding~\cite{Iudice2023islanding} and black-start of a \ac{MG}~\cite{Rokrok2017}. Compared to an average-consensus protocol as used in \cite{Rokrok2017}, min-/max-consensus protocols are used in this work as it does not require fine-tuning of weights based on the network topology for a stable and fast convergence behavior~\cite{Iutzeler2012, Lynch1996}.

\subsection{Contribution and Structure}
We propose a coherent two-step blackout mitigation mechanism that covers both parts of \ac{MG} operation: (2a) \ac{MG} scheduling and (2b) \ac{MG} islanded control. A novel fully distributed \ac{MG} islanded control is designed as a flexibility demand-response schema, using a sequence of max- and min-consensus to coordinate available generation, load, and previously reserved flexibility. An evaluation using a realistic exemplary power grid and hardware-in-the-loop simulation is conducted.

Section~\ref{sec:methodology} describes the proposed methodology, including the flexibility model and the islanded control mechanism. The approach is then evaluated in Section~\ref{sec:eval} and concluded in Section~\ref{sec:concl}.

%% file: sources/02_methodology.tex
\section{Methodology}\label{sec:methodology}
In this paper, \acp{MG} with \( N \) buses in the set \( \mathcal{N} := \{ 1,\dots,N \} \) are operated with a central energy management system in non-blackout, grid-connected state. Different types of software agents are attached to power grid assets at bus \( n \in \mathcal{N} \) and form a \ac{MAS}.
These agents collect and share state information with the energy management system in grid-connected mode and among each other when islanded, in which the availability of central control cannot be guaranteed.
The blackout mitigation is executed for each \( t \) using a scheduling horizon \( \mathcal{T}_{t} := \{ t, t+1,\dots{},t+\mathrm{T} \} \) of \( \mathrm{T} + 1 \) discrete time steps, which sets the desired blackout hold-up time and is limited by the availability of generation and demand forecasts.

\subsection{Agent Types}
\label{sec:methodology_agents}
Four types of flexibility agents can be distinguished. Each represents the respective power grid asset, operates within technical limits, and has an individual social/economic value when operated.

\paragraph{Load Agent (LOAD)}
A load can either be a critical load, which needs to be supplied in any case for social or technical reasons, or an uncritical load, which can be shed. Load shedding is modeled as a binary decision represented by the loads' operational state \(s_{n,\tau} \in \{ 0,1 \} \) for each time \( \tau \in \mathcal{T}_{t} \); 1 means the load is connected and 0 means disconnected. For critical loads \(s_{n,\tau} = 1, \, \forall \tau \in \mathcal{T}_{t} \).
The feasible power steps depend on the intrinsic load \(l^{intrinsic}_{n,\tau} \), i.e., the desirable load, and are given by Eq.~\eqref{eq:load_detach}.
\begin{equation}
     \label{eq:load_detach}
     l_{n,\tau} = l^{intrinsic}_{n,\tau} \cdot{} s_{n,\tau}
\end{equation}
The value of the load agent is the users' willingness to pay for not being shed and, therefore, defines a priority among loads. It translates to load shedding cost \( \mathrm{c^{shed}_{n}} \). Furthermore, repetitively changing the operational state \(s_{n,\tau} \) is non-desireable and comes with switching cost \( \mathrm{c^{sw}_{n}} \) per state change.

\paragraph{Generator Agent (GEN)}
A renewable generation unit is commonly connected by an inverter that allows continuous curtailment. It can provide flexibility by changing its generation setpoint \( g_{n,\tau} \) restricted by its upper generation limit in Eq.~\eqref{eq:generation_limits}.
\begin{equation}\label{eq:generation_limits}
     0 \leq{} g_{n,\tau} \leq{} \overline{g}_{n,\tau}
\end{equation}
A generator might have operational cost \( \mathrm{c^{gen}_{n}} \) per produced kWh. Its value is the opposite of the operating cost and thus negative.

\paragraph{Storage Agent (ESS)}
An \ac{ESS} can store \( f^s_{n,\tau} \) and dispatch \( f^d_{n,\tau} \) power, limited by its minimum and maximum power rates in Eq.~\eqref{eq:flex_power_limits}, and storage capacity in Eq.~\eqref{eq:soc_limits}.
As regular operation at low and high \ac{SoC} harms the expected lifetime of batteries, those limits can be further restricted~\cite{Reddy2011} or be based on the batteries' state of health.
\begin{equation}
     \label{eq:flex_power_limits}
     0 \leq{} f^s_{n,\tau} \leq{} \mathrm{\overline{f}^{s}_{n,\uptau}}
     \quad{}\text{and}\quad{}
     0 \leq{} f^d_{n,\tau} \leq{} \mathrm{\overline{f}^{d}_{n,\uptau}}
\end{equation}
\begin{equation}
     \label{eq:soc_limits}
     \mathrm{\underline{SoC}_{n,\uptau}} \leq{} SoC_{n,\tau} \leq{} \mathrm{\overline{SoC}_{n,\uptau}}
\end{equation}
The time-dependent \ac{SoC} is modeled in Eq.~\eqref{eq:soc_time}, including store efficiency \( \mathrm{\upeta^{s}_{n}} \), dispatch efficiency \( \mathrm{\upeta^{d}_{n}} \), and energy preserving efficiency \( \mathrm{\upeta^{p}_{n}} \). Additional parameters for in- and outflow describe demand-coupled flexibility, such as heat pumps with thermal storage, a secondary heat source (\( \mathrm{in} \)) and heating demand (\( \mathrm{out} \)), or electric vehicles with a driving demand (\( \mathrm{out} \)).
Energy-constrained generators with fuel tanks (\ac{SoC}) are modeled as an \ac{ESS} as well.
\begin{align}
     \label{eq:soc_time}
     SoC_{n,\tau}  = \mathrm{\upeta^{p}_{n}} \cdot{} SoC_{n,\tau-1} + \mathrm{\upeta^s_{n}} \cdot{} f^s_{n,\tau} - \frac{1}{\mathrm{\upeta^d_{n}}} \cdot{} f^d_{n,\tau} + \mathrm{in_{n,\uptau}}-\mathrm{out_{n,\uptau}}
\end{align}
An \ac{ESS} can store energy for later use in islanded mode. Thus, the agents' value is related to its reservation cost \( \mathrm{c^{res}_{n}} \) and usage cost \( \mathrm{c^{use}_{n}} \), which reflects storage degradation or fuel cost.

\paragraph{Grid-forming Resource Agent (GFR)}
A \ac{GFR}, such as a power electronic-based converter with grid-forming capabilities, commonly connected to an \ac{ESS}~\cite{Lasseter2020}, balances out load and generation mismatch, thus provides the required spare flexibility.
Its desired state at the highest value is neither consume (compensate oversupply) nor supply power (compensate undersupply) in order to prolong its operation time by keeping its \ac{SoC} around 50\% to be able to react both to over- and undersupply situations.

\subsection{Two-step Approach}\label{sec:methodology_2_stage}
\begin{figure}[!htb]
     \centering{}
     \includegraphics{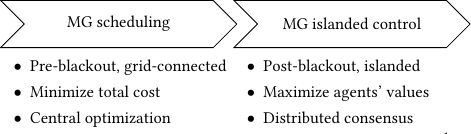}
     \caption{Two-step Approach}\label{fig:two_step_approach}
     \Description{Overview of the two-step approach}
\end{figure}

The two steps of the proposed approach are shown in Figure~\ref{fig:two_step_approach}: (1)~proactive \ac{MG} schedule optimization executed as a \ac{MPC} in grid-connected mode before a blackout event occurs, and (2)~reactive \ac{MG} islanded control to decide on the best flexibility activation iteratively.

As blackouts are rare, only the bare minimum of resources is reserved for later use in critical and high-priority loads. To get the minimum of \( SoC_{n,\tau=t} \), no energy should be left over at the end of the scheduling horizon, and thus \( SoC_{n, \tau=t+\mathrm{T}} = 0, \forall n \in \mathcal{N}\). As objective, the total cost of all agents is minimized, and an optimized schedule for each \ac{ESS} at bus \( n \) is retrieved and defined by \( SoC_{n,\tau}, \forall \tau \in \mathcal{T}_{t} \). The store and dispatch power at time \( t \) to achieve the desired \( SoC_{n,\tau=t} \) is the control variable of the \ac{MPC} to ensure reaching the desired \ac{SoC} in the pre-blackout state. Other types of flexibility must not be reserved but are considered available during the islanded mode. Details of the \ac{MG} scheduling are presented in Appendix \ref{apx:fro}.

Once the \ac{MG} is operated in islanded mode after a blackout, the goal is to maximize the sum of all agent values (inverse cost). This is realized as a distributed demand-response schema with three main objectives: (1) \textit{minimize the balancing power} provided by the \ac{GFR} to prolong its operation, (2) use available generation and reserved flexibility to \textit{serve high-priority loads}, and (3) \textit{store surplus energy} for later use.
In each iteration at time \( t \) (\( \Delta t \) is shorter compared to the \ac{MG} scheduling to also react on non-predictable short-term fluctuations), agents formulate flexibility requests that express a power change with a given value improvement. The highest-valued request is selected by a max-consensus algorithm and matched with the best-fitting flexibility response through a min-consensus round. Both request and response are then activated. The mechanism can be characterized as a distributed, heuristic approach based on well-established consensus algorithms. Agents do not require complete knowledge of the system. Thus, the proposed approach also works in non-perfect situations when some agents are unavailable or communication is interrupted.

\subsection{Consensus-based Control}
The communication network of a \ac{MG} is modeled as an undirected graph \( G=(V, E) \), where \( V \) is the set of agents and \( E \) is the set of links between agents. The existence of the edge \( (i, j) \in E \) represents the ability of the agent \( i \) to exchange information with agent \( j \) bi-directionally. \( (i, i) \notin E, \forall i \in V \), and the graph is connected. \( N_G(i) \) defines the set of neighboring agents of agent \( i \). The communication network could be different from the underlying power grid structure.
All agents share the same system clock, and information exchange occurs in synchronized intervals of \( \Delta t \).
Agents can measure their power \( p^{t} \) and know their desired power state \( p^{o} \). Power values are positive for the consumption of power from the \ac{MG} and negative for the injection of power. \ac{ESS} agents can measure their \ac{SoC} and access their \ac{SoC} schedule.
An agent \( i \) is connected to an asset at bus \( n \in N \). This connection is implicit and thus hidden in the following.

\subsubsection{Flexibility Request}
The flexibility request of an agent \( i \) is formalized as a vector \( \vec{r}_{i}=(r^{p}_{i}, r^{v}_{i}) \), and is defined by the amount of requested power change \( r^{p}_{i}=p^{o} - p^{t} \) and its value \( r^{v}_{i} \). A positive power change means that additional power should be made available, e.g., by increasing generation, while a negative power change represents a need for power reduction. Positive values reflect cost reduction, such as removing shedding costs by connecting a load.

The optimal power state \( p^{o} \) must be within the technical limits of the power grid asset. As summarized in Table~\ref{tab:agents_power}, \( p^{o}=0 \) for \ac{GFR} agents to minimize its balancing power contribution.
LOAD agents have \( l^{intrinsic}_{\tau=t} \) as optimal power state. Generators with \( \mathrm{c^{gen}_{n}} > 0 \) request curtailment to save costs, thus \( p^{o}=0 \).
The \ac{ESS} agent calculates the optimal power state based on its \ac{SoC} schedule. As long as the measured \ac{SoC} is below the schedule, \( p^{o} \) is set to reach the schedule in the next interval. If the measured \ac{SoC} is above the schedule, a charging request without value improvement to charge surplus energy is created.
As the overall system value should be increased, only flexibility requests with \( r^{v}_{i} \ge{} 0 \) are considered.
The value functions are discussed later in Section~\ref{sec:agent_value_functions}.

\begin{table}[!tb]
     \caption{Agents Request and Response Power}
     \label{tab:agents_power}
     \footnotesize{}
     \centering{}
     \begin{tabularx}{\linewidth}{ l l l l }
          \toprule{}
          \multirow{3}{*}{\shortstack[l]{Agent                                                                                                                                                      \\ Type}} & \multicolumn{1}{c}{Request}                           & \multicolumn{2}{c}{Response}                                                                       \\
                                       &                                                       & \( r^{p}>0 \)                                   & \( r^{p}<0 \)                                    \\
                                       & \( r^{p}=p^{o}-p^{t} \)                               & \( a^{p}=p^{t}-\underline{p} \)                 & \( a^{p}=p^{t}-\overline{p} \)                   \\
          \midrule{}
          \vspace{0.3cm}
          \ac{GFR}                     & \( p^{o}=0 \)                                         & \multicolumn{2}{c}{no response}                                                                    \\ \vspace{0.3cm}
          LOAD                         & \( p^{o}={l^{intrinsic}_{t}} \)                       & \( \underline{p}=0 \)                           & \( \overline{p}={l^{intrinsic}_{t}} \)           \\
          \multirow{2}{*}{GEN}         & \multirow{2}{*}{\( p^{o}=0 \)}                        & \( \underline{p}=-\mathrm{\overline{g}_{t}} \)* & \( \overline{p}=0 \)                             \\ \vspace{0.3cm}
                                       &                                                       & \( a^{p}=min(a^{p}, r^{p}) \) ***               & \( a^{p}=max(a^{p}, r^{p}) \) ***                \\
          \multirow{4}{*}{\ac{ESS} **} & \( p^{o}=\frac{SoC_{t+1} - SoC_{t}}{\Delta t} \)      & \( \underline{p}=max(p^{o}, 0.0) \)             & \( \overline{p}=\mathrm{\overline{f}^{s}_{t}} \) \\
                                       & if \( r^{p} < 0 \):                                   & \( a^{p}=min(a^{p}, r^{p}) \) ***               & \( a^{p}=max(a^{p}, r^{p}) \) ***                \\
                                       & \( \quad r^{p}=\mathrm{\overline{f}^{s}_{t}}-p^{t} \) &                                                 &                                                  \\
                                       & \( \quad r^{v}=0 \)                                   &                                                 &                                                  \\
          \bottomrule{}
     \end{tabularx}
     \\ * The system is seen as consuming system, thus GEN limit is negative; ** Only if \( \mathrm{\underline{SoC}_{t}} \leq{} SoC_{t} \leq{} \mathrm{\overline{SoC}_{t}} \); *** Continuously controllable power resources scale their response power down to the requested power.
\end{table}

A max-consensus round is conducted to find the highest-valued request (pseudocode in Appendix~\ref{apx:consensus_algos}). Each agent \( i \) initializes its maximum known request \( \vec{r}_{i\_max} \) with its own request \( \vec{r}_{i} \), if it is above a given threshold. To eventually reach consensus, every agent \( i \) sends its maximum known request \( \vec{r}_{i\_max} \) to its direct neighbors \( N_G(i) \). Whenever an agent \( i \) receives a request from a neighboring agent \( j \), it updates its maximum known request \( \vec{r}_{i\_max} \) with the received \( \vec{r}_{j} \), if the received flexibility value \( r^{v}_{j} \) is larger than the value of \( \vec{r}_{i\_max} \). The update is then communicated to the neighbors until no further updates are sent. In the case of value equality, a predefined unique priority is used, e.g., distributed together with the \ac{SoC} schedules or from an initial setup of the \ac{MG}.
Convergence is ensured at the latest after \( diam(G) \) rounds. A precalculated iteration limit could be set in the agents as the upper limit of the diameter (\( iter^{max} > diam(G) \))~\cite{Lynch1996}.

\subsubsection{Flexibility Response}
Once the highest-valued flexibility request is obtained, other agents apart from the request-winning agent can provide a response based on their available flexibility. The response of an agent \( i \) is formalized with a flexibility vector \( \vec{a}_{i}=(a^{p}_{i},a^{v}_{i}) \), in which \( a^{p}_{i} \) corresponds to the offered power change at value \( a^{v}_{i} \).
As the response should meet the request, a positive power request \( r^{p}_{i}>0 \) can only be answered with a positive power response \( a^{p}_{i}=p^{t}-\underline{p_{i}} \) and a negative power request only with a negative power response \( a^{p}_{i}=p^{t}-\overline{p_{i}} \). The response is skipped if the response power is below a given threshold. Furthermore, responses with a higher value loss than the value gained from the request are filtered out.
GEN and LOAD agents offer a power change up to their technical limits. \ac{ESS} agents offer charging up to the maximizing charging rate and discharging up to a calculated optimal charging rate that ensures not to fall below the \ac{SoC} schedule. GEN and \ac{ESS} agents can scale their power response continuously. Thus, their power change response is limited to the requested power for a perfect fit to the request. \ac{GFR} agents do not provide a response as they automatically balance out unmatched supply and demand as part of their grid-forming function.

The min-consensus round to find the best-fitting response is similar to the max-consensus. Instead of the maximum flexibility value, the smallest Manhattan distance between the request and response vectors is searched for. This ensures that the best-fitting response of the non-comparable dimensions of power and value is selected. The algorithm is detailed in Appendix~\ref{apx:consensus_algos}.

\subsubsection{Flexibility Activation}
After running both consensus rounds, the flexibility request with the highest value and the best-fitting response are found. If no response exists, that is, no flexibility is available or all responses are more expensive than the request value, no activation is performed. GEN and \ac{ESS} agents can be controlled with a continuous power setpoint, therefore, their requested power is reduced to the best-fitting response to maintain power balance.
Both the request and the response are activated simultaneously to force a quick power balance in the \ac{MG}. Therefore, the \ac{GFR} only needs to react during ramping delays.
In some cases, request and response power will not perfectly match, for example, when a smaller load is disconnected for a larger, higher-priority load. This situation would lead to a long-term power imbalance, further formulated as a flexibility request by the \ac{GFR} agent in the next iteration. LOAD agents that have changed their state will suspend further participation for a given time to avoid oscillation.

\subsubsection{Agents Flexibility Value}\label{sec:agent_value_functions}
\begin{figure}[!tb]
     \centering{}
     \includegraphics{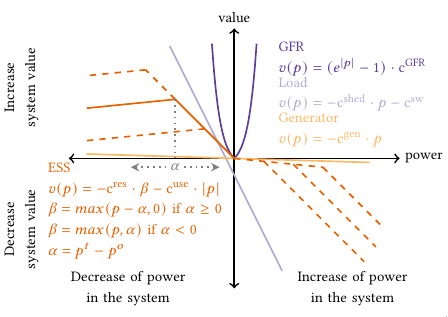}
     \caption{Value Functions for Different Agent Types}
     \label{fig:dcs_value_functions}
     \Description{Value Functions for Different Agent Types.}
\end{figure}

The value functions vary by agent type and reflect the cost structure for providing flexibility. The time factor \( \Delta t \) is neglected as only a relative value difference between agents must be achieved. Different approaches to derive value functions have been investigated by simulations. Figure \ref{fig:dcs_value_functions} plots the value functions per agent type in the vector space.
The functions return the value for flexibility activation. For a flexibility request that defines the required power change so that its activation is compensated, the negative requested power change needs to be inserted in the value function to obtain the value of its activation.

The most valuable asset is the \ac{GFR}, thus the value of its request is an exponential function of its absolute loading. In both under and oversupply situations, flexibility is requested to zero out power contribution by the \ac{GFR}. The value function of GEN and LOAD reflect the linear costs with \( \mathrm{c^{gen}_{n}} \) per power change for generators and \( \mathrm{c^{shed}_{n}} \) per power change and \( \mathrm{c^{sw}_{n}} \) per activation for loads.
\ac{ESS} agents value their flexibility activation based on their \ac{SoC} linear to \( \mathrm{c^{res}} \) and \( \mathrm{c^{use}} \) per power change. In this way, charging and discharging toward the scheduled \ac{SoC} level is benefited, and moving away is penalized. If its \ac{SoC} is above the schedule, only \( \mathrm{c^{use}} \) is considered and oversupply can be absorbed and dispatched when requested.

%% file: sources/03_evaluation.tex
\section{Evaluation}\label{sec:eval}
The evaluation is focused on the applicability and effectiveness of the consensus-based approach (\ac{MG} scheduling is evaluated in Appendix~\ref{apx:fro_eval}). Two timescales are interesting: (1) the blackstart phase with an ordered reconnection of loads and (2) the whole prediction horizon to evaluate if all critical loads are served as planned.

\subsection{Simulation Setup}
\begin{figure}[!htb]
    \centering{}
    \includegraphics[width=0.85\linewidth]{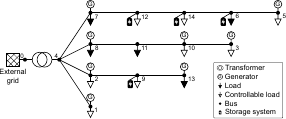}
    \caption{SimBench Grid \texttt{``1-LV-rural1--1-sw''}}
    \label{fig:simbench_grid}
    \Description{Overview of selected SimBench power grid.}
\end{figure}

A representative 13-bus low-voltage rural power grid is selected from the SimBench dataset~\cite{Meinecke2020} and shown in Figure~\ref{fig:simbench_grid}. It comes with demand and generation profiles in a 15-minute resolution projected for 2024. 14 loads, 8 photovoltaic systems with 468.2~kWp, and four \ac{ESS} with 311.5~kWh capacity are present. It is a net-generating grid with a yearly consumption of 201.9~MWh and a yearly generation of 302.3~MWh. \acp{ESS} have a C-rate of 0.5, meaning their full capacity can be (dis) charged in two hours.
Loads with a peak power lower than 3 kW are considered non-controllable (critical) loads, resulting in 8 out of 14 controllable loads. Costs are user inputs and have been fixed for this exemplary evaluation: \( \mathrm{c^{shed}} \) are randomly chosen within [0,1], \( \mathrm{c^{res}} \) are 0.1 (bus 12), 0.2 (bus 9), 0.3 (bus 14) and 0.4 (bus 6), and \( \mathrm{c^{use}} \) are 0.001, 0.002, 0.003, 0.004 respectively. \( \mathrm{c^{sw}} \) is set to 0.0001 for all loads, and \( \mathrm{c^{gen}} \) is set to zero, as all generation units are photovoltaic systems. The wall clock time \( t \) is set to 2\textsuperscript{nd} August 00:00.
The threshold for agents to state a flexibility request or response is set to 0.5~kW. The quarter-hourly profiles are interpolated to a one-minute resolution using the quadratic method. Loads do not offer flexibility requests or responses for the next 15~minutes after a state change to avoid oscillation.

\begin{figure}[!tb]
    \centering{}
    \includegraphics{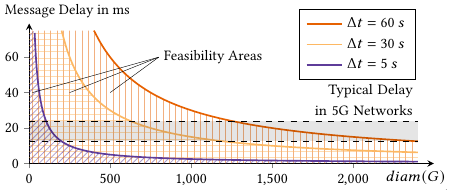}
    \caption{Feasible Delay and Graph Diameter for \( \Delta t \) with typically delays in 5G networks~\cite{Christopoulou2021}}
    \label{fig:dcs_delays}
    \Description{Impact of message delay and communication network diameter on the interval time of the \ac{MG} islanded control.}
\end{figure}

The maximum number of consecutive communication messages to run one iteration, which is one max- and one min-consensus round, is defined by \( 2 \times diam(G) \)~\cite{Lynch1996}. There exists an upper limit for the execution interval \( \Delta t \) that depends on \( diam(G) \) and the expected message delay. As long as these parameters are within the feasibility area, both consensus rounds can be guaranteed to be finished within the corresponding \( \Delta t \). Different settings are compared in Figure~\ref{fig:dcs_delays}. \( \Delta t = 60~s\) is more than enough to drive a comparable large \ac{MG} even with a poor communication network and therefore used here. Therefore, message delay and detailed communication topology modeling are neglected. Furthermore, this interval avoids high-frequent setpoint changes.

\subsection{Blackstart}
\begin{figure}[!b]
    \centering{}
    \begin{subfigure}[b]{0.49\linewidth}
        \centering{}
        \includegraphics[width=\linewidth]{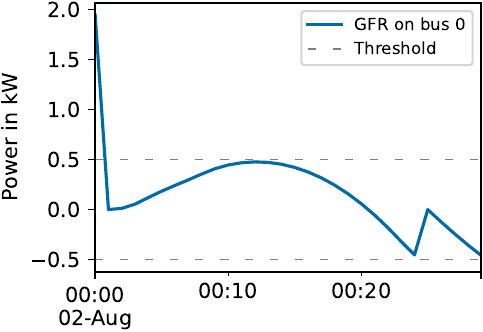}
        \caption{\ac{GFR} Loading}
        \label{fig:dcs_gfr_loading_blackstart}
    \end{subfigure}
    \hfill{}
    \begin{subfigure}[b]{0.49\linewidth}
        \centering{}
        \includegraphics[width=\linewidth]{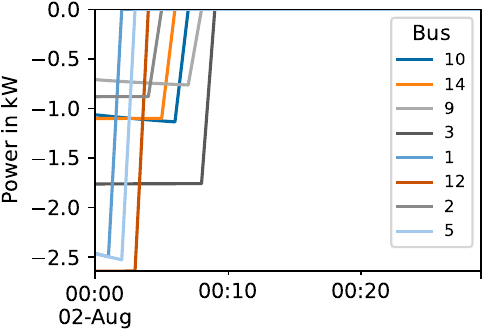}
        \caption{Load Blackstart}
        \label{fig:dcs_loading_blackstart}
    \end{subfigure}
    \caption{Blackstart}
    \label{fig:dcs_blackstart}
    \Description{Initial blackstart of the \ac{MG} showing the iterative connection of additional loads.}
\end{figure}

The initial blackstart is shown in Figure~\ref{fig:dcs_blackstart} and is based on an over-reserved \ac{MG} schedule with a confidence level of 0.95 for the chance constraints on forecasts (compare Appendix~\ref{apx:fro_eval_conservative}).
First, the \ac{GFR} starts to supply around 2~kW power to all non-controllable loads within the grid. The agents then start and formulate their flexibility requests. Controllable loads, which are off by default, have a request to connect and the \ac{GFR} agent has a request to reduce its power supply. The highest-valued flexibility request by the \ac{GFR} is supplied with the cheapest \ac{ESS} at bus 12. One by one, all loads are connected (see Figure~\ref{fig:dcs_loading_blackstart}) and served by the same over-reserved \ac{ESS}. At the same time, the loading of \ac{GFR} is kept below the threshold of \(\pm{}\)0.5~kW. After nine iterations, all loads are connected and the \ac{GFR} does not need to provide balancing power. The blackstart time could be shortened by reducing \( \Delta t\).

\subsection{Long-term Run}
\subsubsection{Oversupply}
\begin{figure}[!tb]
    \centering{}
    \begin{subfigure}[b]{0.49\linewidth}
        \centering{}
        \includegraphics[width=\linewidth]{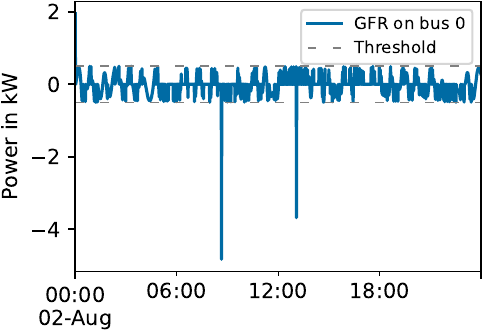}
        \caption{\ac{GFR} Loading}
        \label{fig:dcs_os_gfr_loading}
    \end{subfigure}
    \hfill{}
    \begin{subfigure}[b]{0.49\linewidth}
        \centering{}
        \includegraphics[width=\linewidth]{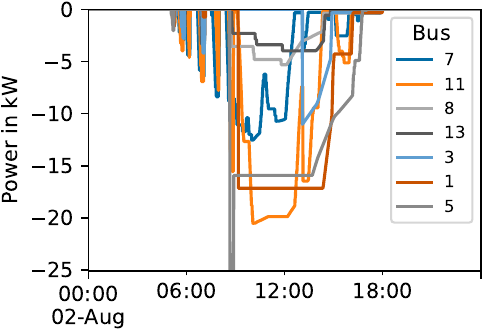}
        \caption{Generation Curtailment}
        \label{fig:dcs_os_generation_curtailment}
    \end{subfigure}
    \begin{subfigure}[b]{0.49\linewidth}
        \centering{}
        \includegraphics[width=\linewidth]{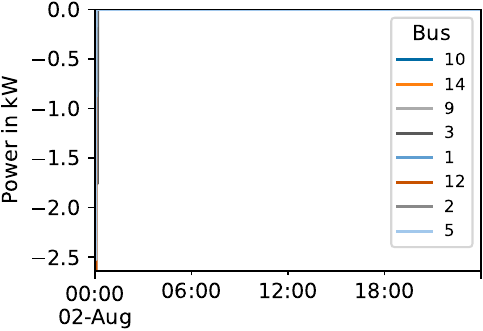}
        \caption{Load Shedding}
        \label{fig:dcs_os_load_shedding}
    \end{subfigure}
    \begin{subfigure}[b]{0.49\linewidth}
        \centering{}
        \includegraphics[width=\linewidth]{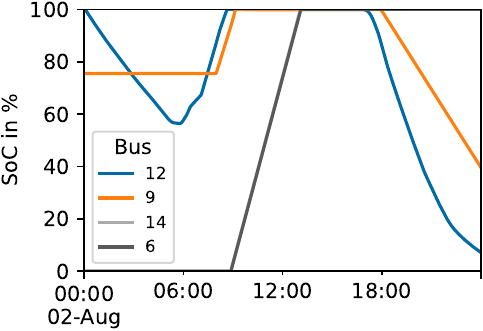}
        \caption{Storage Usage}
        \label{fig:dcs_os_storage_usage}
    \end{subfigure}
    \caption{Long-term (Oversupply)}
    \label{fig:dcs_os_longterm}
    \Description{Run of the \ac{MG} islanded control for a full day, showing that all loads are served with previously reserved energy.}
\end{figure}

Because \ac{ESS} reservation has been made on a conservative forecast with a confidence level of 0.95, over-reserved energy is available for all loads during the whole execution horizon. Therefore, loads are only briefly disconnected during blackstart (Figure~\ref{fig:dcs_os_load_shedding}).
Furthermore, all \acp{ESS} are fully charged with the energy of the photovoltaic system at 13:05 and only the two cheaper \acp{ESS} are required to supply power for loads until the end of the blackout hold-up time (Figure~\ref{fig:dcs_os_storage_usage}).
Oversupply from the photovoltaic system during the day that can not be stored in the \acp{ESS} is curtailed as visualized in Figure~\ref{fig:dcs_os_generation_curtailment}.
The loading of the \ac{GFR} stays within its limits most of the time, depicted in Figure~\ref{fig:dcs_os_gfr_loading}. Short spikes at 08:39 and 13:05 are caused by a sudden stop of charging from \acp{ESS} at buses 12, 14, and 6. A ramp rate limit could reduce those spikes.
The \ac{GFR} has a mean loading of -21.56~W (consumption exceeds feed-in) and would require only 0.75~kWh of energy storage to compensate for fluctuations during the execution horizon.

\subsubsection{Undersupply}
The forecasted profiles are applied in the \ac{MG} scheduling without a safety margin for forecast uncertainty. Furthermore, the biggest and cheapest \ac{ESS} at bus 12 is removed, reducing the total storage capacity by 47\% and requiring all \acp{ESS} to be fully used (see Appendix~\ref{apx:fro_eval_undersupply}). When applying the actual profiles in the \ac{MG} islanded control simulation, this scenario delivers undersupply situations, which need to be handled.

\begin{figure}[!tb]
    \centering{}
    \begin{subfigure}[b]{0.49\linewidth}
        \centering{}
        \includegraphics[width=\linewidth]{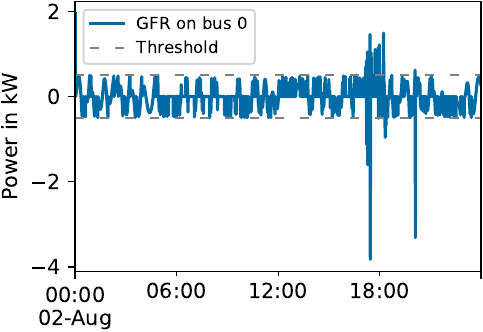}
        \caption{\ac{GFR} Loading}
        \label{fig:dcs_us_gfr_loading}
    \end{subfigure}
    \hfill{}
    \begin{subfigure}[b]{0.49\linewidth}
        \centering{}
        \includegraphics[width=\linewidth]{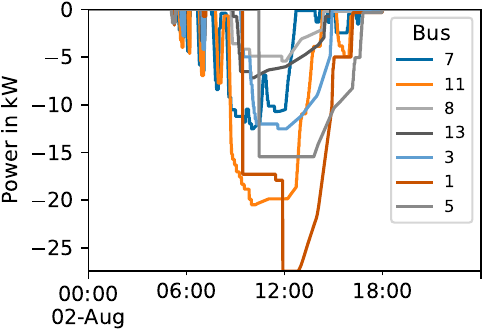}
        \caption{Generation Curtailment}
        \label{fig:dcs_us_generation_curtailment}
    \end{subfigure}
    \begin{subfigure}[b]{0.49\linewidth}
        \centering{}
        \includegraphics[width=\linewidth]{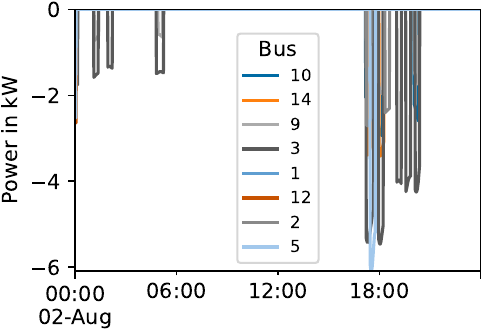}
        \caption{Load Shedding}
        \label{fig:dcs_us_load_shedding}
    \end{subfigure}
    \begin{subfigure}[b]{0.49\linewidth}
        \centering{}
        \includegraphics[width=\linewidth]{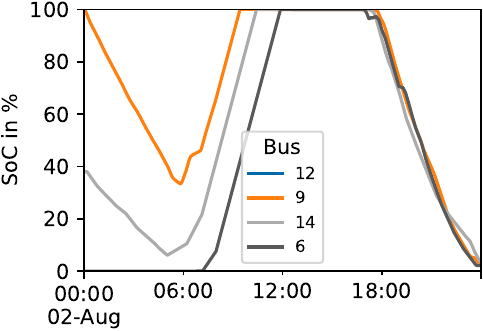}
        \caption{Storage Usage}
        \label{fig:dcs_us_storage_usage}
    \end{subfigure}
    \caption{Long-term (Undersupply)}
    \label{fig:dcs_us_longterm}
    \Description{Run of the \ac{MG} islanded control for a full day, showing that all loads are served with previously reserved energy.}
\end{figure}

The initial blackstart is the same as with the first run. Instead of keeping the load at bus 3 disconnected as optimized in the \ac{MG} scheduling, the \ac{MG} islanded control tries to use minor spare energy to pick it up (Figure~\ref{fig:dcs_us_load_shedding}). Unfortunately, the stored energy is not enough to drive all loads during the first night and loads need to be disconnected temporarily to keep the \ac{SoC} above the schedule. After sunset, loads have been underpredicted and are partially shut down but reconnected once the \ac{SoC} is above the schedule.
Other than forecasted, more energy is available from generation starting from 6:00 and the reserved energy is not fully used (Figure~\ref{fig:dcs_us_storage_usage}).
Generation curtailment during the day (Figure~\ref{fig:dcs_us_generation_curtailment}) and \ac{GFR} loading (Figure~\ref{fig:dcs_us_gfr_loading}) are similar to the first run.
In total, 18.55~kWh of load have been shed compared to 30.29~kWh as forecasted in the \ac{MG} scheduling. All critical non-controllable loads have been supplied during the execution horizon.

\subsection{Hardware-in-the-Loop Simulation}
The \ac{MG} islanded control is further applied in a hardware-in-the-loop simulation. The setup consists of a TyphoonHIL 604 for power grid and power electronics emulation and a smart grid converter controller board, which runs a direct voltage control algorithm with virtual impedance-based grid-forming capabilities, similar to~\cite{Zhang2010}. To ensure traceability, the power grid model is reduced to one \ac{GFR} (34.5~kWp), one photovoltaic system (6~kWp, \( c^{gen}=0 \)), one \ac{ESS} (5~kWh, C-rate of 1, no reservation with \( p^{o}=-5 kW \), \( c^{res}=0.1 \), \( c^{use}=0.2 \)), one critical load (3~kW) and one controllable load (7~kW, \( c^{shed}=1.0 \), \( c^{sw}=0.1 \)). The photovoltaic system and the \ac{ESS} have a ramp rate of 0.5~p.u./s, whereas loads are switched immediately. \( \Delta t \) is set to 5~s and generation and demand is fixed over time.

\begin{figure}[!tb]
    \centering{}
    \includegraphics[width=\linewidth]{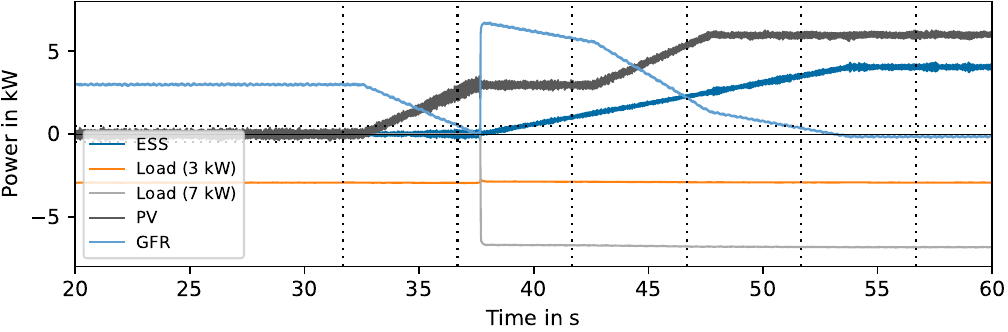}
    \caption{Power Generation (Positive) and Consumption (Negative)}
    \label{fig:hil_results}
    \Description{\ac{HIL} execution of \ac{MG} islanded control showing the power and frequency over time.}
\end{figure}

As visualized in Figure~\ref{fig:hil_results}, the initial contribution of the \ac{GFR} until 33~s after blackstart is 3~kW to supply the critical load. Then, the agents execute the first out of 6 control rounds. The individual request and response vectors for each round are plotted in Appendix~\ref{apx:hil_vectors}.
For the first execution, the highest-valued request is an undersupply notice from the \ac{GFR}, which is responded by ramping up the photovoltaic system.
In the second execution, the initial power unbalance is settled, and the controllable load has the highest value. Its pickup is responded by ramping up the \ac{ESS}, favored over the photovoltaic system because it can provide 6~kW with small usage costs instead of only remaining 3~kW.
As the best-fitting response from the photovoltaic system is not instantly provided, the \ac{GFR} compensates the power mismatch and requests a balance in the third execution. The best-fitting response is from the photovoltaic system. Now, both the photovoltaic system and the \ac{ESS} increase their power supply for a fast reaction.
In the fourth execution round, the power unbalance at the \ac{GFR} is not yet solved, but the photovoltaic system is approaching its limit. Thus, the \ac{ESS} is selected again as the best-fitting response and its setpoint is updated to the current needs.
All following execution rounds only have a charging request from the \ac{ESS}, which can not be answered. Due to the update of the \ac{ESS} setpoint in iteration four, an overshoot of the target and thus oversupply has been prevented.

%% file: sources/04_conclusions.tex
\section{Conclusions}\label{sec:concl}
This paper proposes a two-step blackout prevention approach capable of prolonging \ac{MG} islanded operation time during a wide-area power outage. The proactive flexibility reservation, performed before a blackout event, minimizes the required storage reserve so that the remaining capacity can be used for market participation. During a blackout, the heuristic flexibility demand response algorithm minimizes the balancing power provided by grid-forming resources, supplies high-priority loads, and stores surplus energy from generators. Hardware-in-the-loop and long-term simulations showed the designed algorithm's applicability and convergence. The execution of the \ac{MG} islanded control during the predefined blackout hold-up time showed that the three objectives were fulfilled. Even in bad forecast situations with an undersupply, 40\% less energy is shed than planned in the \ac{MG} scheduling. The hardware-in-the-loop simulation further showed that with iterative execution of the \ac{MG} islanded control, setpoints are continuously updated to achieve a fast convergence towards the optimal situation and accurate handling of ramp rate limits.

During an emergency scenario, such as a wide-area blackout, a communication system can experience degradation and even loss of service. Although the area of a \ac{MG} is usually relatively small, and a limited amount of access network components is required to provide communication between the agents, the impact of communication service disruption, such as loss of communicating agents, increased message delay and loss factors, on the performance of the developed methods will be observed in future work.

%% file: sources/appendix.tex
\appendix{}

\section{\acs{MG} Scheduling}\label{apx:fro}
The \ac{MG} scheduling problem to optimize \ac{ESS} reservation and resulting \ac{SoC} schedule is detailed in the following. This optimization is embedded in a \ac{MPC} with store and dispatch power of a \ac{ESS} at time \( t \) as the control variable to eventually reach the desired reservation state \( SoC_{n,\tau=t} \).

\subsection{Problem Formulation}
To reserve the bare minimum for operating the \ac{MG} in island mode within the given prediction horizon \( \mathcal{T}_{t} \), the sum of all agent costs and power flow costs is minimized.
First, generation cost \( C_{GEN} \) summarizes individual cost per produced amount of energy in Eq.~\eqref{eq:cost_gen}, which can be neglected for renewable generators.
\( C_{ESS} \) combines the total amount of reserved \ac{SoC} at time \( t \) with storage usage cost in Eq.~\eqref{eq:cost_ess}. The reservation cost may relate to missed benefits from self-consumption or market participation, and the usage cost, such as cyclic battery aging, is considered with a linear model similar to~\cite{Shi2019, Guo2018}.
\( C_{LOAD} \) considers load shedding and load switching cost in Eq.~\eqref{eq:cost_load}.
Finally, the cost of the power flows \( C_{pf} \) provides an optimal distribution of the storage reserves closest to the supply-demand mismatch as part of the power flow cost in Eq.~\eqref{eq:cost_pf}. This also reduces the impact of power losses as part of the DC power flow approximation.
\begin{align}
    \label{eq:cost_gen} C_{GEN} = & \sum^{N}_{n=1}\sum^{t+T}_{\tau=t} \mathrm{c^{gen}_n} \cdot{} g_{n,\tau} \cdot{} \Delta{\tau}                                              \\
    \begin{split}
        \label{eq:cost_ess}
        C_{ESS} =      & \sum^{N}_{n=1}\biggl(\mathrm{c^{res}_n} \cdot{} SoC_{n,\tau=t}\\
        & + \sum^{t+T}_{\tau=t} \mathrm{c^{use}_n} \cdot{} (\mathrm{\upeta^s_{n}} \cdot{} f^s_{n,\tau} + \frac{1}{\mathrm{\upeta^d_{n}}} \cdot{} f^d_{n,\tau}) \cdot{} \Delta{\tau} \biggl)
    \end{split}                         \\
    \begin{split}
        \label{eq:cost_load}
        C_{LOAD} =    & \sum^{N}_{n=1}\sum^{T}_{\tau=t} \biggl(\mathrm{c^{shed}_n} \cdot{} ({l^{intrinsic}_{n,\tau}}-l_{n,\tau}) \cdot{} \Delta{\tau}\\
        & + \mathrm{c^{sw}_n} \cdot{} \big| s_{n,\tau}-s_{n,\tau-1} \big| \biggl)
    \end{split} \\
    \label{eq:cost_pf} C_{pf} =   & \sum^{t+T}_{\tau=t}\sum^{N}_{n=1}\sum^{N}_{m > n} \mathrm{c^{flow}} \cdot{} \big|f_{n,m,\tau}\big|
\end{align}

To get the minimum of \( SoC_{n,\tau=t} \) for each \ac{ESS} to be reserved for critical and high-priority loads, no energy should be left over at the end of the scheduling horizon as in Eq.~\eqref{eq:flex_soc_final}. Further, it is not required to prohibit simultaneous storing and dispatching as their sum is minimized in Eq.~\eqref{eq:cost_ess}.
\begin{equation}
    \label{eq:flex_soc_final}
    SoC_{n, \tau=t+T} = 0
\end{equation}

Power flow constraints are given by the nodal power balance according to Kirchhoff's current law in Eq.~\eqref{eq:nodal_power_balance}, linearized branch flow calculation based on branch susceptance in Eq.~\eqref{eq:susceptance} and flow limits to prevent overloading of power lines in Eq.~\eqref{eq:branch_flow_limits}.
\begin{align}
    \sum_{m} f_{n,m,\tau}  & = g_{n,\tau}-l_{n,\tau}+f^d_{n,\tau}-f^s_{n,\tau} \label{eq:nodal_power_balance} \\
    f_{n,m,\tau}           & = \mathrm{B_{n,m}} (\theta_{n,\tau}-\theta_{m,\tau}) \label{eq:susceptance}      \\
    \big|f_{n,m,\tau}\big| & \leq{} \mathrm{\overline{f}_{n,m}} \label{eq:branch_flow_limits}
\end{align}

The scheduling problem is formulated as an optimal power flow problem by
\begin{align}
    \label{eq:problem}
    \begin{split}
        \underset{f^s_{n,\tau}, f^d_{n,\tau}, g_{n,\tau}, s_{n,\tau}}{\min} & C_{GEN} + C_{ESS} + C_{LOAD} + C_{pf}                               \\
        \text{s.t. }                                              & \text{\eqref{eq:load_detach}~--~\eqref{eq:branch_flow_limits}}~.
    \end{split}
\end{align}

\subsection{Parametric Chance Constraints}
As the upper limit of generation forecast \( \overline{g}_{n,\tau} \) is uncertain, Eq.~\eqref{eq:generation_limits} is replaced by a chance constraint in Eq.~\eqref{eq:unc_gen} in which \( \widehat{g}_{n,\tau} \) represents the uncertain generation and \( \gamma{} = 1 - \epsilon{} \) defines the confidence level to meet the constraint (i.e., \( \epsilon{} \) describes the risk of not meeting the constraint).
Similarly, intrinsic load \( {l^{intrinsic}_{n,\tau}} \) is subject to uncertainty. As \( {l^{intrinsic}_{n,\tau}} \) is minimized as part of agent costs in Eq.~\eqref{eq:cost_load} and~\eqref{eq:load_detach}, an uncertain load forecast variable \( \widehat{l}_{n,\tau} \) is added as a lower bound to the intrinsic load in Eq.~\eqref{eq:unc_load} to be fulfilled with confidence~\( \gamma{} \).
\begin{align}
    \label{eq:unc_gen}
    \mathbb{P} \{ \underline{g}_{n,\tau} \leq{} g_{n,\tau} \leq{} \widehat{g}_{n,\tau} \} & \geq{} \gamma{} = 1-\epsilon{} \\
    \label{eq:unc_load}
    \mathbb{P} \{ l^{intrinsic}_{n,\tau} \geq{} \widehat{l}_{n,\tau} \}                   & \geq{} \gamma{} = 1-\epsilon{}
\end{align}

Those probabilistic individual chance constraints are not in a closed form as required by standard solvers. For the sake of demonstration of the general approach and even if not perfectly representing reality, the forecast errors of generation and load are assumed to be independent Gaussian distributed random variables with \( \mathcal{N} (\upmu, \upsigma^2) \)~\cite{Hemmati2020, Antoniadou-Plytaria2022, Ciftci2019}.
This allows us to convert inequalities with an uncertain parameter to a deterministic linear constraint by calculating the \(\epsilon \)-quantile of the normal distribution (that is, the inverse of the cumulative density function \( \Phi \)). Therefore, Eq.~\eqref{eq:unc_gen} and~\eqref{eq:unc_load} are reformulated in Eq.~\eqref{eq:unc_gen_relaxed} and~\eqref{eq:unc_load_relaxed}.
\begin{align}
    \label{eq:unc_gen_relaxed}
    g^{\min}_{n,\tau} \leq{} g_{n,\tau} & \leq{} \mathrm{\widehat{g}_{n,\uptau}}-\mathrm{\upmu^g_{n,\uptau}}-\mathrm{\upsigma^g_{n,\uptau}} \cdot{} \Phi^{-1} (\gamma{})     \\
    \label{eq:unc_load_relaxed}
    l^{intrinsic}_{n,\tau}              & \geq{} \mathrm{\widehat{l}_{n,\uptau}} + \mathrm{\upmu^l_{n,\uptau}} + \mathrm{\upsigma^l_{n,\uptau}} \cdot{} \Phi^{-1} (\gamma{})
\end{align}
As a result, the chance-constrained \ac{MG} scheduling problem is a mixed-integer linear problem, which is efficiently solvable by standard solvers, and formulated as follows:
\begin{align}
    \label{eq:problem_cc}
    \begin{split}
        \underset{f^s_{n,\tau}, f^d_{n,\tau}, g_{n,\tau}, s_{n,\tau}}{\min}  &  C_{GEN} + C_{ESS} + C_{LOAD} + C_{pf}  \\
        \text{s.t. }                                              & \text{\eqref{eq:load_detach},~\eqref{eq:flex_power_limits}~--~\eqref{eq:branch_flow_limits},~\eqref{eq:unc_gen_relaxed},~\eqref{eq:unc_load_relaxed}}
    \end{split}
\end{align}

\subsection{Evaluation}\label{apx:fro_eval}
The evaluation is applied on the exemplary power grid as defined in Section~\ref{sec:eval}. The focus here is on the impact of the confidence level on the reserved \ac{ESS} capacity.

\begin{figure}[!tb]
    \centering{}
    \begin{subfigure}[b]{0.49\linewidth}
        \centering{}
        \includegraphics[width=\linewidth]{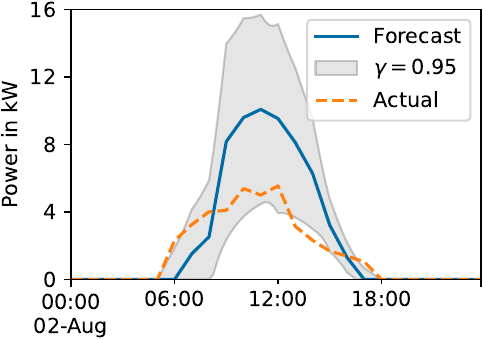}
        \caption{Photovoltaic system at bus 8}
        \label{fig:gen_forecast}
    \end{subfigure}
    \hfill{}
    \begin{subfigure}[b]{0.49\linewidth}
        \centering{}
        \includegraphics[width=\linewidth]{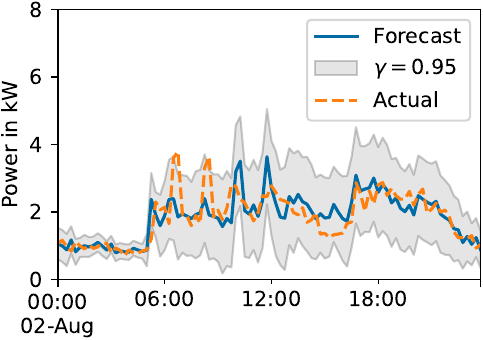}
        \caption{Load at bus 10}
        \label{fig:load_forecast}
    \end{subfigure}
    \caption{Same-as-Yesterday Forecast with Error range}
    \label{fig:forecast}
    \Description{Exemplary same-as-yesterday forecast of one photovoltaic system and one load for one day.}
\end{figure}

Power flow cost \( \mathrm{c^{flow}} \) is set to 0.0001 to have only a slight impact on the overall optimization result but still minimize power flow. The prediction horizon is set to 24~hours and a same-as-yesterday forecast model is used. The forecasted power is compared with the actual power for each quarter-hour of the day over one year, resulting in 365 error samples per interval of the day.
The error distribution is almost normal and a pair of \( (\upmu, \upsigma) \) can be fitted per interval.
The forecast of two exemplary assets with a confidence interval of 0.95 is plotted in Figure~\ref{fig:forecast}.
More advanced forecasting techniques are out-of-scope for this paper.

\subsubsection{Impact of Confidence Level}
\begin{figure}[!b]
    \centering{}
    \includegraphics[width=\linewidth]{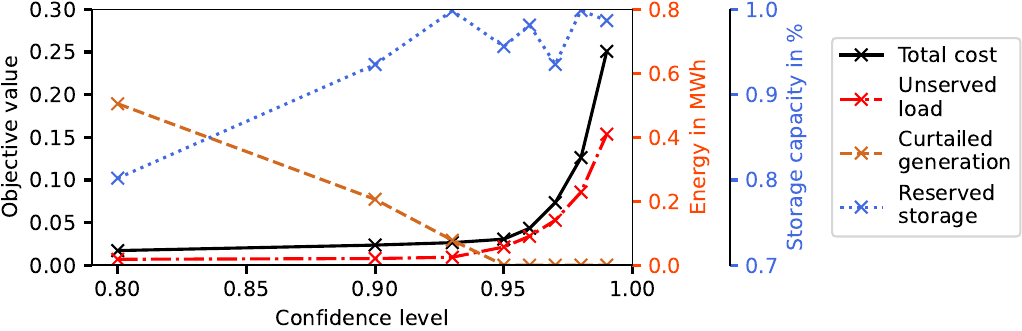}
    \caption{Impact of Confidence Level on the \ac{MG} Scheduling}
    \label{fig:fro_impact_confidence}
    \Description{\ac{FRO} with different confidence levels for the chance constraints.}
\end{figure}

The confidence level within the chance constraints can be seen as a control of conservatism. However, this conservatism comes with a price. The total cost (=objective value) increases with a higher confidence level because a worse situation is considered (less generation, more consumption). As shown in Figure~\ref{fig:fro_impact_confidence}, the amount of unserved load and, therefore, the total cost increases rapidly once all self-generated energy is used and most of the battery capacity is reserved. Compared to total costs of 0.008 for a perfect forecast, costs triple at 0.9 confidence (0.024), almost quadruple at 0.95 confidence (0.031), and multiply by 31 at 0.99 confidence (0.251).

\subsubsection{Conservative Reservation}\label{apx:fro_eval_conservative}
\begin{figure}[!htb]
    \centering{}
    \begin{subfigure}[b]{0.49\linewidth}
        \centering{}
        \includegraphics[width=\linewidth]{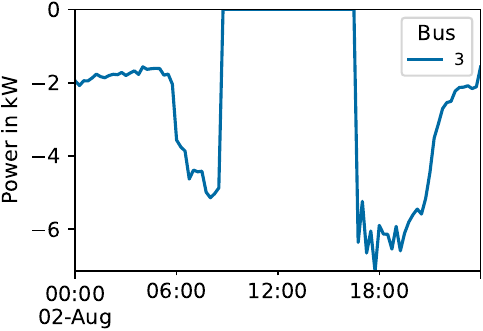}
        \caption{Load Shedding}
        \label{fig:load_shedding}
    \end{subfigure}
    \hfill{}
    \begin{subfigure}[b]{0.49\linewidth}
        \centering{}
        \includegraphics[width=\linewidth]{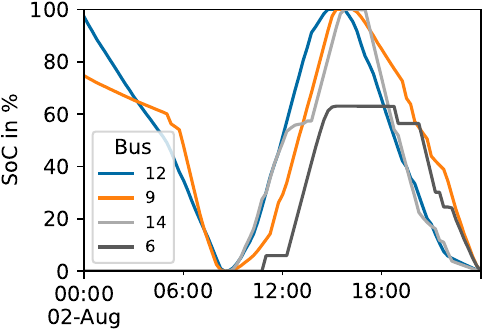}
        \caption{Storage Usage}
        \label{fig:storage_usage}
    \end{subfigure}
    \caption{\ac{MG} Schedule with Confidence Level 0.95}
    \label{fig:fro_run}
    \Description{Optimized schedules of loads and \ac{ESS} as results of the \ac{FRO}.}
\end{figure}

Running the optimization with 0.95 confidence level results in all loads being served except for the load on bus 3. The shedding cost for this load is lower than the storage reservation cost, and, therefore, the load is turned off during the night, as shown in Figure~\ref{fig:load_shedding}. It can still be served as long as free energy is produced by the photovoltaic systems in daylight.
From the \ac{SoC} levels in Figure~\ref{fig:storage_usage}, it can be observed that the cheaper \ac{ESS} on bus 12 (100\%) and bus 9 (75.5\%) are reserved for the first night. During the day, all \acp{ESS} are charged from self-produced energy to supply high-priority loads in the second night.

The mean absolute DC approximation error (in particular, the neglected power loss), measured by the contribution of the slack bus when applying the optimized schedules in an AC power flow, is 527~W, which translates to 2.4\% of the total generated energy during the scheduling horizon.
Power loss is negligible compared to the mean absolute forecast error of the whole grid with 34.71~kW.

\subsubsection{Undersupply Scenario}\label{apx:fro_eval_undersupply}
\begin{figure}[!htb]
    \centering{}
    \begin{subfigure}[b]{0.49\linewidth}
        \centering{}
        \includegraphics[width=\linewidth]{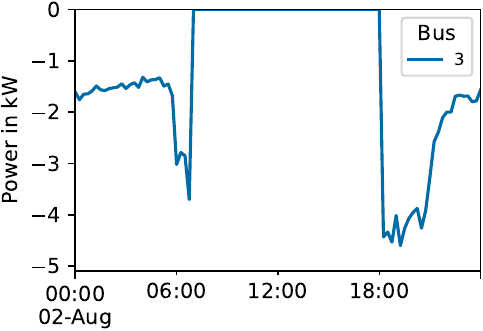}
        \caption{Load Shedding}
    \end{subfigure}
    \hfill{}
    \begin{subfigure}[b]{0.49\linewidth}
        \centering{}
        \includegraphics[width=\linewidth]{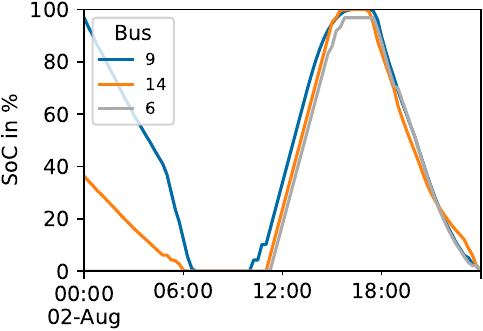}
        \caption{Storage Usage}
    \end{subfigure}
    \begin{subfigure}[b]{0.49\linewidth}
        \centering{}
        \includegraphics[width=\linewidth]{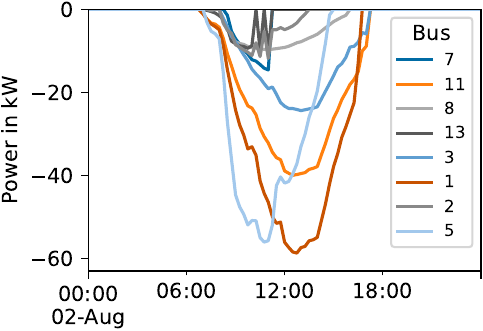}
        \caption{Generation Curtailment}
    \end{subfigure}
    \caption{\ac{MG} Scheduling with Directly Applied Forecasts in Undersupply Scenario}
    \label{fig:fro_run_undersupply}
    \Description{Optimized schedules of loads and \ac{ESS} as results of the \ac{FRO} in undersupply scenario.}
\end{figure}

The forecasted profiles are applied in the scheduling problem in Eq.~\eqref{eq:problem} without considering forecast error uncertainty. Furthermore, the biggest and cheapest \ac{ESS} at bus 12 is disconnected, reducing the total storage capacity by 47\%. The results of the schedule optimization are plotted in Figure~\ref{fig:fro_run_undersupply}. The load at bus 3 cannot be supplied during the night and the available \ac{ESS} capacity is insufficient to charge all surplus energy during the day and later supply the load at bus 3.

\section{Consensus Algorithms}\label{apx:consensus_algos}
The max-consensus algorithm as pseudocode to find the highest-valued flexibility request is shown in Algorithm~\ref{alg:max_consensus}.
\begin{algorithm}[!htb]
    \caption{Flexibility Request: Max-Consensus}\label{alg:max_consensus}
    \DontPrintSemicolon{}
    \LinesNotNumbered{}
    \textbf{Ensure} \( iter^{max} > diam(G), N_G(i) \neq \varnothing \) \;
    \( r^{p}_{i} \gets p^{o}_{i} - p^{t}_{i} \) \tcp*{Create flexibility request}
    \uIf{\( r^{p}_{i} < \mathrm{{p}^{threshold}} \)}{
        \( r^{p}_{i} \gets 0 \) \tcp*{Skip if power below threshold}
    }
    \( r^{v}_{i} \gets v(-r^{p}_{i})\) \tcp*{Calculate value of flexibility}
    \( \vec{r}_{i} \gets (r^{p}_{i},r^{v}_{i}) \) \;
    \( \vec{r}_{i\_max} \gets \vec{r}_{i} \) \tcp*{Initialize max known request}
    \textcolor{cb_4}{
        \For{\( iter^{max} \)}{
            \textcolor{cb_3}{
                \ForEach{\( j \in N_G(i) \)}{
                    \textcolor{cb_2}{
                        \( send(j, \vec{r}_{i\_max}) \) \tcp*{Send to agent \( j \)}
                    }
                    \textcolor{cb_1}{
                        \( \vec{r}_{j} \gets receive(j) \) \tcp*{Receive from agent \( j \)}
                        \uIf{\( \vec{r}^{v}_{j} > \vec{r}^{v}_{i\_max} \)}{
                            \( \vec{r}_{i\_max} \gets \vec{r}_{j} \)
                        }
                    }
                }
            }
        }
    }
\end{algorithm}

\begin{figure}[!htb]
    \centering{}
    \includegraphics{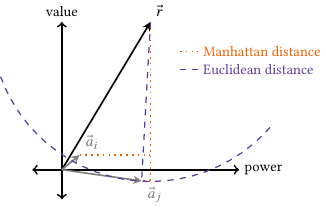}
    \caption{Request and Response Vectors}
    \label{fig:dcs_vectors}
    \Description{A flexibility request vector and two response vectors that compare Euclidean with Manhattan distance.}
\end{figure}

The min-consensus algorithm to find the best-fitting flexibility response is shown in Algorithm~\ref{alg:min_consensus}.
We use the Manhattan distance, not the Euclidean distance, because the two dimensions of the flexibility vectors have different scales and are therefore not comparable. As can be seen in the example in Figure~\ref{fig:dcs_vectors}, the Manhattan distance favors \( \vec{a}_{j} \) instead of \( \vec{a}_{i} \), which has a smaller Euclidean distance but provides only a fifth of the requested power with only small value improvement. Another option would be to normalize the response vector to the request vector to overcome this issue.

\begin{algorithm}[!htb]
    \DontPrintSemicolon{}
    \SetNoFillComment{}
    \LinesNotNumbered{}
    \SetKwProg{Fn}{Function}{ is}{end}

    \caption{Flexibility Response: Min-Consensus}\label{alg:min_consensus}

    \textbf{Ensure} \( iter^{max} > diam(G), N_G(i) \neq \varnothing \) \;

    \Fn{\( d(\vec{r}, \vec{a}) \)}{
    return \( |{r^{p}-a^{p}}| + |{r^{v} - a^{v}}| \) \;
    }

    \tcc{Create flexibility response}
    \( \vec{a}_{i} \gets (a^{p}_{i},a^{v}_{i}) \) \tcp*{According to Table~\ref{tab:agents_power}}
    \uIf{\( a^{p}_{i} < \mathrm{{p}^{threshold}} \)}{
        \( a^{p}_{i} \gets 0 \) \tcp*{Skip if power below threshold}
    }
    \uIf{\( {SoC} \leq \mathrm{\underline{SoC}} \) \textbf{or} \( {SoC} \geq \mathrm{\overline{SoC}} \)}{
        \( a^{p}_{i} \gets 0 \) \tcp*{Skip if not feasible (only for \ac{ESS})}
    }
    \uIf{\( \vec{r}^{v}_{i\_max} - a^{v}_{i} < 0 \)}{
        \( a^{p}_{i} \gets 0 \) \tcp*{Skip if response is too expensive}
    }
    \( \vec{a}_{i\_min} \gets \vec{a}_{i} \) \tcp*{Initialize min response}
    \textcolor{cb_4}{
        \For{\( iter^{max} \)}{
            \textcolor{cb_3}{
                \ForEach{\( j \in N_G(i) \)}{
                    \textcolor{cb_2}{
                        \( send(j, \vec{a}_{i\_min}) \) \tcp*{Send to agent \( j \)}
                    }
                    \textcolor{cb_1}{
                        \( \vec{a}_{j} \gets receive(j) \) \tcp*{Receive from agent \( j \)}
                        \uIf{\(d(\vec{r}_{i\_max}, \vec{a}_{j})  < d(\vec{r}_{i\_max}, \vec{a}_{i\_min}) \)}{
                            \( \vec{a}_{i\_min} \gets \vec{a}_{j} \) \;
                        }
                    }
                }
            }
        }
    }
\end{algorithm}

\section{Hardware-in-the-Loop Request and Response Vectors}\label{apx:hil_vectors}
\begin{figure}[!htb]
    \centering{}
    \begin{subfigure}[b]{0.49\linewidth}
        \centering{}
        \includegraphics[width=\linewidth]{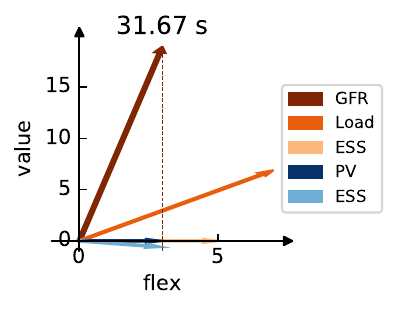}
    \end{subfigure}
    \begin{subfigure}[b]{0.49\linewidth}
        \centering{}
        \includegraphics[width=\linewidth]{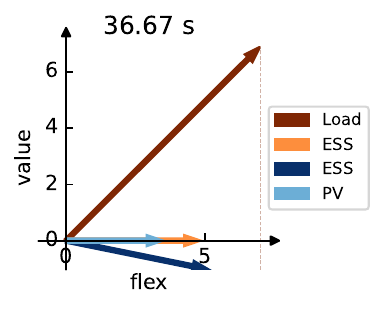}
    \end{subfigure}
    \begin{subfigure}[b]{0.49\linewidth}
        \centering{}
        \includegraphics[width=\linewidth]{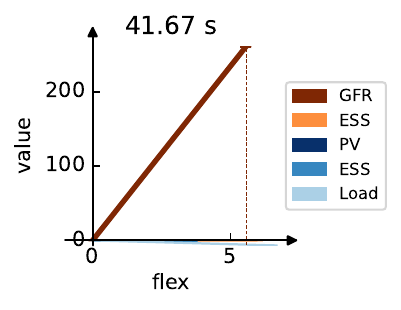}
    \end{subfigure}
    \begin{subfigure}[b]{0.49\linewidth}
        \centering{}
        \includegraphics[width=\linewidth]{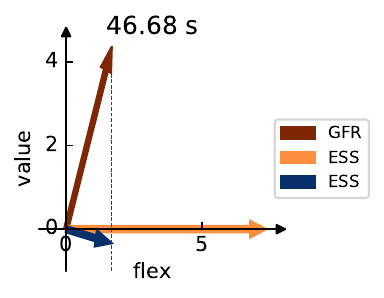}
    \end{subfigure}
    \begin{subfigure}[b]{0.49\linewidth}
        \centering{}
        \includegraphics[width=\linewidth]{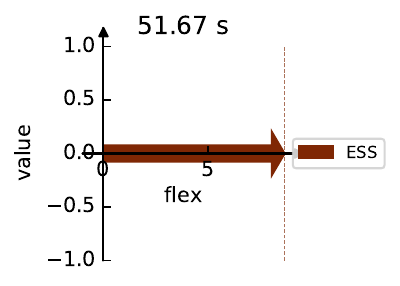}
    \end{subfigure}
    \begin{subfigure}[b]{0.49\linewidth}
        \centering{}
        \includegraphics[width=\linewidth]{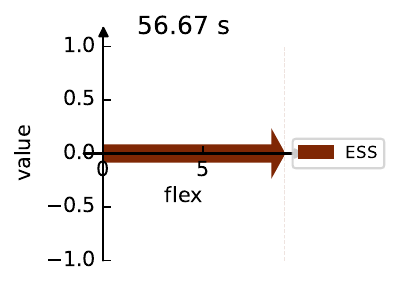}
    \end{subfigure}
    \caption{Hardware-in-the-Loop Flexibility Request (Orange/Red) and Response (Blue) Vectors}
    \label{fig:hil_vectors}
    \Description{Hardware-in-the-loop execution of \ac{MG} islanded control showing the power and frequency over time.}
\end{figure}

The individual iterations of the hardware-in-the-loop simulation are shown in Figure~\ref{fig:hil_vectors}, in which orange (lower) to red (higher) represent flexibility requests ordered by request value, and light blue (less-fitting) to dark blue (best-fitting) represent all flexibility responses ordered by their minimum Manhattan distance to the flexibility request.

%% file: arxiv.bbl

\begin{thebibliography}{37}


\ifx \showCODEN    \undefined \def \showCODEN     #1{\unskip}     \fi
\ifx \showDOI      \undefined \def \showDOI       #1{#1}\fi
\ifx \showISBNx    \undefined \def \showISBNx     #1{\unskip}     \fi
\ifx \showISBNxiii \undefined \def \showISBNxiii  #1{\unskip}     \fi
\ifx \showISSN     \undefined \def \showISSN      #1{\unskip}     \fi
\ifx \showLCCN     \undefined \def \showLCCN      #1{\unskip}     \fi
\ifx \shownote     \undefined \def \shownote      #1{#1}          \fi
\ifx \showarticletitle \undefined \def \showarticletitle #1{#1}   \fi
\ifx \showURL      \undefined \def \showURL       {\relax}        \fi
\providecommand\bibfield[2]{#2}
\providecommand\bibinfo[2]{#2}
\providecommand\natexlab[1]{#1}
\providecommand\showeprint[2][]{arXiv:#2}

\bibitem[Anderson and Suryanarayanan(2020)]%
        {Anderson2020}
\bibfield{author}{\bibinfo{person}{Alexander~A. Anderson} {and} \bibinfo{person}{Siddharth Suryanarayanan}.} \bibinfo{year}{2020}\natexlab{}.
\newblock \showarticletitle{{Review of Energy Management and Planning of Islanded Microgrids}}.
\newblock \bibinfo{journal}{\emph{CSEE Journal of Power and Energy Systems}} \bibinfo{volume}{6}, \bibinfo{number}{2} (\bibinfo{year}{2020}), \bibinfo{pages}{329--343}.
\newblock
\showISSN{20960042}
\urldef\tempurl%
\url{https://doi.org/10.17775/CSEEJPES.2019.01080}
\showDOI{\tempurl}


\bibitem[Antoniadou-Plytaria et~al\mbox{.}(2022)]%
        {Antoniadou-Plytaria2022}
\bibfield{author}{\bibinfo{person}{Kyriaki Antoniadou-Plytaria}, \bibinfo{person}{David Steen}, \bibinfo{person}{Le~Anh Tuan}, \bibinfo{person}{Ola Carlson}, \bibinfo{person}{Baraa Mohandes}, {and} \bibinfo{person}{Mohammad Ali~Fotouhi Ghazvini}.} \bibinfo{year}{2022}\natexlab{}.
\newblock \showarticletitle{{Scenario-based Stochastic Optimization for Energy and Flexibility Dispatch of a Microgrid}}.
\newblock \bibinfo{journal}{\emph{IEEE Transactions on Smart Grid}} \bibinfo{volume}{13}, \bibinfo{number}{5} (\bibinfo{year}{2022}), \bibinfo{pages}{3328--3341}.
\newblock
\showISSN{19493061}
\urldef\tempurl%
\url{https://doi.org/10.1109/TSG.2022.3175418}
\showDOI{\tempurl}


\bibitem[Christopoulou et~al\mbox{.}(2021)]%
        {Christopoulou2021}
\bibfield{author}{\bibinfo{person}{Maria Christopoulou}, \bibinfo{person}{Georgios Xilouris}, \bibinfo{person}{Athanasios Sarlas}, \bibinfo{person}{Harilaos Koumaras}, \bibinfo{person}{Michail~Alexandros Kourtis}, {and} \bibinfo{person}{Themistoklis Anagnostopoulos}.} \bibinfo{year}{2021}\natexlab{}.
\newblock \showarticletitle{{5G Experimentation: The Experience of the Athens 5GENESIS Facility}}. In \bibinfo{booktitle}{\emph{2021 IFIP/IEEE International Symposium on Integrated Network Management (IM)}}. \bibinfo{publisher}{IEEE}, \bibinfo{address}{Bordeaux, France}, \bibinfo{pages}{783--787}.
\newblock


\bibitem[Ciftci et~al\mbox{.}(2019)]%
        {Ciftci2019}
\bibfield{author}{\bibinfo{person}{Okan Ciftci}, \bibinfo{person}{Mahdi Mehrtash}, \bibinfo{person}{Farnaz Safdarian}, {and} \bibinfo{person}{Amin Kargarian}.} \bibinfo{year}{2019}\natexlab{}.
\newblock \showarticletitle{{Chance-Constrained Microgrid Energy Management with Flexibility Constraints Provided by Battery Storage}}. In \bibinfo{booktitle}{\emph{2019 IEEE Texas Power and Energy Conference (TPEC)}}. \bibinfo{publisher}{IEEE}, \bibinfo{address}{College Station, TX, USA}, \bibinfo{pages}{1--6}.
\newblock
\showISBNx{978-1-5386-9284-4}
\urldef\tempurl%
\url{https://doi.org/10.1109/TPEC.2019.8662200}
\showDOI{\tempurl}


\bibitem[Danner and de~Meer(2022)]%
        {Danner2022}
\bibfield{author}{\bibinfo{person}{Dominik Danner} {and} \bibinfo{person}{Hermann de Meer}.} \bibinfo{year}{2022}\natexlab{}.
\newblock \showarticletitle{{Max-consensus protocol to determine the regulated node in distributed voltage regulation}}.
\newblock \bibinfo{journal}{\emph{Energy Informatics}} \bibinfo{volume}{5}, \bibinfo{number}{1} (\bibinfo{year}{2022}), \bibinfo{pages}{1--12}.
\newblock
\showISSN{25208942}
\urldef\tempurl%
\url{https://doi.org/10.1186/s42162-022-00211-w}
\showDOI{\tempurl}


\bibitem[Deplano et~al\mbox{.}(2023)]%
        {Deplano2023}
\bibfield{author}{\bibinfo{person}{Diego Deplano}, \bibinfo{person}{Mauro Franceschelli}, {and} \bibinfo{person}{Alessandro Giua}.} \bibinfo{year}{2023}\natexlab{}.
\newblock \showarticletitle{{Dynamic Min and Max Consensus and Size Estimation of Anonymous Multiagent Networks}}.
\newblock \bibinfo{journal}{\emph{IEEE Trans. Automat. Control}} \bibinfo{volume}{68}, \bibinfo{number}{1} (\bibinfo{date}{jan} \bibinfo{year}{2023}), \bibinfo{pages}{202--213}.
\newblock
\showISBNx{2000035000}
\showISSN{0018-9286}
\urldef\tempurl%
\url{https://doi.org/10.1109/TAC.2021.3135452}
\showDOI{\tempurl}


\bibitem[Espina et~al\mbox{.}(2020)]%
        {espina2020controlstrategies}
\bibfield{author}{\bibinfo{person}{Enrique Espina}, \bibinfo{person}{Jacqueline Llanos}, \bibinfo{person}{Claudio Burgos-Mellado}, \bibinfo{person}{Roberto Cárdenas-Dobson}, \bibinfo{person}{Manuel Martínez-Gómez}, {and} \bibinfo{person}{Doris Sáez}.} \bibinfo{year}{2020}\natexlab{}.
\newblock \showarticletitle{Distributed Control Strategies for Microgrids: An Overview}.
\newblock \bibinfo{journal}{\emph{IEEE Access}}  \bibinfo{volume}{8} (\bibinfo{year}{2020}), \bibinfo{pages}{193412--193448}.
\newblock
\urldef\tempurl%
\url{https://doi.org/10.1109/ACCESS.2020.3032378}
\showDOI{\tempurl}


\bibitem[Fotis et~al\mbox{.}(2022)]%
        {Fotis2022}
\bibfield{author}{\bibinfo{person}{Georgios Fotis}, \bibinfo{person}{Vasiliki Vita}, {and} \bibinfo{person}{Theodoros~I. Maris}.} \bibinfo{year}{2022}\natexlab{}.
\newblock \showarticletitle{{Risks in the European Transmission System and a Novel Restoration Strategy for a Power System after a Major Blackout}}.
\newblock \bibinfo{journal}{\emph{Applied Sciences}} \bibinfo{volume}{13}, \bibinfo{number}{1} (\bibinfo{date}{dec} \bibinfo{year}{2022}), \bibinfo{pages}{83}.
\newblock
\showISSN{2076-3417}
\urldef\tempurl%
\url{https://doi.org/10.3390/app13010083}
\showDOI{\tempurl}


\bibitem[Giannini et~al\mbox{.}(2016)]%
        {Giannini2016}
\bibfield{author}{\bibinfo{person}{Silvia Giannini}, \bibinfo{person}{Antonio Petitti}, \bibinfo{person}{Donato {Di Paola}}, {and} \bibinfo{person}{Alessandro Rizzo}.} \bibinfo{year}{2016}\natexlab{}.
\newblock \showarticletitle{{Asynchronous Max-Consensus Protocol with Time Delays: Convergence Results and Applications}}.
\newblock \bibinfo{journal}{\emph{IEEE Transactions on Circuits and Systems I: Regular Papers}} \bibinfo{volume}{63}, \bibinfo{number}{2} (\bibinfo{year}{2016}), \bibinfo{pages}{256--264}.
\newblock
\showISSN{15580806}
\urldef\tempurl%
\url{https://doi.org/10.1109/TCSI.2015.2512721}
\showDOI{\tempurl}


\bibitem[Guo and Zhao(2018)]%
        {Guo2018}
\bibfield{author}{\bibinfo{person}{Yuanxiong Guo} {and} \bibinfo{person}{Chaoyue Zhao}.} \bibinfo{year}{2018}\natexlab{}.
\newblock \showarticletitle{{Islanding-aware robust energy management for microgrids}}.
\newblock \bibinfo{journal}{\emph{IEEE Transactions on Smart Grid}} \bibinfo{volume}{9}, \bibinfo{number}{2} (\bibinfo{year}{2018}), \bibinfo{pages}{1301--1309}.
\newblock
\showISSN{19493053}
\urldef\tempurl%
\url{https://doi.org/10.1109/TSG.2016.2585092}
\showDOI{\tempurl}


\bibitem[{Haes Alhelou} et~al\mbox{.}(2019)]%
        {HaesAlhelou2019}
\bibfield{author}{\bibinfo{person}{Hassan {Haes Alhelou}}, \bibinfo{person}{Mohamad Hamedani-Golshan}, \bibinfo{person}{Takawira Njenda}, {and} \bibinfo{person}{Pierluigi Siano}.} \bibinfo{year}{2019}\natexlab{}.
\newblock \showarticletitle{{A Survey on Power System Blackout and Cascading Events: Research Motivations and Challenges}}.
\newblock \bibinfo{journal}{\emph{Energies}} \bibinfo{volume}{12}, \bibinfo{number}{4} (\bibinfo{date}{feb} \bibinfo{year}{2019}), \bibinfo{pages}{682}.
\newblock
\showISBNx{8415683111}
\showISSN{1996-1073}
\urldef\tempurl%
\url{https://doi.org/10.3390/en12040682}
\showDOI{\tempurl}


\bibitem[Hemmati et~al\mbox{.}(2020)]%
        {Hemmati2020}
\bibfield{author}{\bibinfo{person}{Mohammad Hemmati}, \bibinfo{person}{Behnam Mohammadi-Ivatloo}, \bibinfo{person}{Mehdi Abapour}, {and} \bibinfo{person}{Amjad Anvari-Moghaddam}.} \bibinfo{year}{2020}\natexlab{}.
\newblock \showarticletitle{{Optimal Chance-Constrained Scheduling of Reconfigurable Microgrids Considering Islanding Operation Constraints}}.
\newblock \bibinfo{journal}{\emph{IEEE Systems Journal}} \bibinfo{volume}{14}, \bibinfo{number}{4} (\bibinfo{year}{2020}), \bibinfo{pages}{5340--5349}.
\newblock
\showISSN{19379234}
\urldef\tempurl%
\url{https://doi.org/10.1109/JSYST.2020.2964637}
\showDOI{\tempurl}


\bibitem[Iudice et~al\mbox{.}(2023)]%
        {Iudice2023islanding}
\bibfield{author}{\bibinfo{person}{Francesco~Lo Iudice}, \bibinfo{person}{Ricardo Cardona-Rivera}, \bibinfo{person}{Antonio Grotta}, \bibinfo{person}{Marco Coraggio}, {and} \bibinfo{person}{Mario di Bernardo}.} \bibinfo{year}{2023}\natexlab{}.
\newblock \showarticletitle{{Consensus-Based Distributed Intentional Controlled Islanding of Power Grids}}.
\newblock \bibinfo{journal}{\emph{IEEE Transactions on Control of Network Systems}}  \bibinfo{volume}{10} (\bibinfo{year}{2023}), \bibinfo{pages}{1--11}.
\newblock
\showISSN{2325-5870}
\urldef\tempurl%
\url{https://doi.org/10.1109/TCNS.2023.3277805}
\showDOI{\tempurl}


\bibitem[Iutzeler et~al\mbox{.}(2012)]%
        {Iutzeler2012}
\bibfield{author}{\bibinfo{person}{Franck Iutzeler}, \bibinfo{person}{Philippe Ciblat}, {and} \bibinfo{person}{J{\'{e}}r{\'{e}}mie Jakubowicz}.} \bibinfo{year}{2012}\natexlab{}.
\newblock \showarticletitle{{Analysis of Max-Consensus Algorithms in Wireless Channels}}.
\newblock \bibinfo{journal}{\emph{IEEE Transactions on Signal Processing}} \bibinfo{volume}{60}, \bibinfo{number}{11} (\bibinfo{date}{11} \bibinfo{year}{2012}), \bibinfo{pages}{6103--6107}.
\newblock
\showISSN{1053-587X}
\urldef\tempurl%
\url{https://doi.org/10.1109/TSP.2012.2211593}
\showDOI{\tempurl}


\bibitem[Lasseter et~al\mbox{.}(2020)]%
        {Lasseter2020}
\bibfield{author}{\bibinfo{person}{Robert~H. Lasseter}, \bibinfo{person}{Zhe Chen}, {and} \bibinfo{person}{Dinesh Pattabiraman}.} \bibinfo{year}{2020}\natexlab{}.
\newblock \showarticletitle{{Grid-Forming Inverters: A Critical Asset for the Power Grid}}.
\newblock \bibinfo{journal}{\emph{IEEE Journal of Emerging and Selected Topics in Power Electronics}} \bibinfo{volume}{8}, \bibinfo{number}{2} (\bibinfo{year}{2020}), \bibinfo{pages}{925--935}.
\newblock
\showISSN{21686785}
\urldef\tempurl%
\url{https://doi.org/10.1109/JESTPE.2019.2959271}
\showDOI{\tempurl}


\bibitem[Lin et~al\mbox{.}(2007)]%
        {Lin2007}
\bibfield{author}{\bibinfo{person}{Jie Lin}, \bibinfo{person}{A~Stephen Morse}, {and} \bibinfo{person}{Brian~DO Anderson}.} \bibinfo{year}{2007}\natexlab{}.
\newblock \showarticletitle{{The Multi-Agent Rendezvous Problem. Part 2: The Asynchronous Case}}.
\newblock \bibinfo{journal}{\emph{SIAM Journal on Control and Optimization}} \bibinfo{volume}{46}, \bibinfo{number}{6} (\bibinfo{year}{2007}), \bibinfo{pages}{2120--2147}.
\newblock
\urldef\tempurl%
\url{https://doi.org/10.1137/040620564}
\showDOI{\tempurl}


\bibitem[Lu et~al\mbox{.}(2022)]%
        {Lu2022}
\bibfield{author}{\bibinfo{person}{Jinghang Lu}, \bibinfo{person}{Xiaojie Liu}, \bibinfo{person}{Xiaochao Hou}, {and} \bibinfo{person}{Peng Wang}.} \bibinfo{year}{2022}\natexlab{}.
\newblock \showarticletitle{{A Distributed Control Strategy for State-of-Charge Balance of Energy Storage without Continuous Communication in AC Microgrids}}.
\newblock \bibinfo{journal}{\emph{IEEE Transactions on Sustainable Energy}} \bibinfo{volume}{14}, \bibinfo{number}{1} (\bibinfo{year}{2022}), \bibinfo{pages}{206--216}.
\newblock
\showISSN{19493037}
\urldef\tempurl%
\url{https://doi.org/10.1109/TSTE.2022.3206327}
\showDOI{\tempurl}


\bibitem[Lynch(1996)]%
        {Lynch1996}
\bibfield{author}{\bibinfo{person}{Nancy~A. Lynch}.} \bibinfo{year}{1996}\natexlab{}.
\newblock \bibinfo{booktitle}{\emph{Distributed Algorithms}}.
\newblock \bibinfo{publisher}{Morgan Kaufmann Publishers Inc.}, \bibinfo{address}{San Francisco, CA, USA}.
\newblock
\showISBNx{9780080504704}


\bibitem[Mahmood and Blaabjerg(2022)]%
        {Mahmood2022}
\bibfield{author}{\bibinfo{person}{Hisham Mahmood} {and} \bibinfo{person}{Frede Blaabjerg}.} \bibinfo{year}{2022}\natexlab{}.
\newblock \showarticletitle{{Autonomous Power Management of Distributed Energy Storage Systems in Islanded Microgrids}}.
\newblock \bibinfo{journal}{\emph{IEEE Transactions on Sustainable Energy}} \bibinfo{volume}{13}, \bibinfo{number}{3} (\bibinfo{year}{2022}), \bibinfo{pages}{1507--1522}.
\newblock
\showISSN{19493037}
\urldef\tempurl%
\url{https://doi.org/10.1109/TSTE.2022.3156393}
\showDOI{\tempurl}


\bibitem[Mahzarnia et~al\mbox{.}(2020)]%
        {Mahzarnia2020}
\bibfield{author}{\bibinfo{person}{Maedeh Mahzarnia}, \bibinfo{person}{Mohsen~Parsa Moghaddam}, \bibinfo{person}{Payam~Teimourzadeh Baboli}, {and} \bibinfo{person}{Pierluigi Siano}.} \bibinfo{year}{2020}\natexlab{}.
\newblock \showarticletitle{{A Review of the Measures to Enhance Power Systems Resilience}}.
\newblock \bibinfo{journal}{\emph{IEEE Systems Journal}} \bibinfo{volume}{14}, \bibinfo{number}{3} (\bibinfo{date}{sep} \bibinfo{year}{2020}), \bibinfo{pages}{4059--4070}.
\newblock
\showISSN{1932-8184}
\urldef\tempurl%
\url{https://doi.org/10.1109/JSYST.2020.2965993}
\showDOI{\tempurl}


\bibitem[{Manuel Mauricio} et~al\mbox{.}(2021)]%
        {ManuelMauricio2021}
\bibfield{author}{\bibinfo{person}{Juan {Manuel Mauricio}}, \bibinfo{person}{Kyriaki-Nefeli Malamaki}, \bibinfo{person}{Jose {Maria Maza-Ortega}}, \bibinfo{person}{Georgios~C. Kryonidis}, \bibinfo{person}{Manuel {Barragan- Villarejo}}, \bibinfo{person}{Spyros~I. Gkavanoudis}, {and} \bibinfo{person}{Charis~S. Demoulias}.} \bibinfo{year}{2021}\natexlab{}.
\newblock \showarticletitle{{Short-term Energy Recovery Control for Virtual Inertia Provision by Renewable Energy Sources}}. In \bibinfo{booktitle}{\emph{2021 IEEE 30th International Symposium on Industrial Electronics (ISIE)}}, Vol.~\bibinfo{volume}{2021-June}. \bibinfo{publisher}{IEEE}, \bibinfo{address}{Kyoto, Japan}, \bibinfo{pages}{1--6}.
\newblock
\showISBNx{978-1-7281-9023-5}
\urldef\tempurl%
\url{https://doi.org/10.1109/ISIE45552.2021.9576213}
\showDOI{\tempurl}


\bibitem[Meinecke et~al\mbox{.}(2020)]%
        {Meinecke2020}
\bibfield{author}{\bibinfo{person}{Steffen Meinecke}, \bibinfo{person}{D{\v{z}}anan Sarajli{\'{c}}}, \bibinfo{person}{Simon~Ruben Drauz}, \bibinfo{person}{Annika Klettke}, \bibinfo{person}{Lars~Peter Lauven}, \bibinfo{person}{Christian Rehtanz}, \bibinfo{person}{Albert Moser}, {and} \bibinfo{person}{Martin Braun}.} \bibinfo{year}{2020}\natexlab{}.
\newblock \showarticletitle{{SimBench - A benchmark dataset of electric power systems to compare innovative solutions based on power flow analysis}}.
\newblock \bibinfo{journal}{\emph{Energies}} \bibinfo{volume}{13}, \bibinfo{number}{12} (\bibinfo{year}{2020}), \bibinfo{numpages}{19}~pages.
\newblock
\showISSN{1996-1073}
\urldef\tempurl%
\url{https://doi.org/10.3390/en13123290}
\showDOI{\tempurl}


\bibitem[Monteiro and Peixoto(2020)]%
        {Monteiro2020}
\bibfield{author}{\bibinfo{person}{João~C. Monteiro} {and} \bibinfo{person}{Alessandro~Jacoud Peixoto}.} \bibinfo{year}{2020}\natexlab{}.
\newblock \showarticletitle{Convergence and Stability Properties of a Dynamic Maximum Consensus Estimator}.
\newblock \bibinfo{journal}{\emph{IFAC-PapersOnLine}} \bibinfo{volume}{53}, \bibinfo{number}{2} (\bibinfo{year}{2020}), \bibinfo{pages}{2885--2890}.
\newblock
\showISSN{2405-8963}
\urldef\tempurl%
\url{https://doi.org/10.1016/j.ifacol.2020.12.960}
\showDOI{\tempurl}
\newblock
\shownote{21st IFAC World Congress}.


\bibitem[Nejad et~al\mbox{.}(2009)]%
        {Nejad2009}
\bibfield{author}{\bibinfo{person}{Behrang~Monajemi Nejad}, \bibinfo{person}{Sid~Ahmed Attia}, {and} \bibinfo{person}{J{\"{o}}rg Raisch}.} \bibinfo{year}{2009}\natexlab{}.
\newblock \showarticletitle{{Max-consensus in a max-plus algebraic setting: The case of fixed communication topologies}}. In \bibinfo{booktitle}{\emph{2009 XXII International Symposium on Information, Communication and Automation Technologies}}. \bibinfo{publisher}{IEEE}, \bibinfo{address}{Sarajevo, Bosnia and Herzegovina}, \bibinfo{numpages}{7}~pages.
\newblock
\showISBNx{9781424442218}
\urldef\tempurl%
\url{https://doi.org/10.1109/ICAT.2009.5348437}
\showDOI{\tempurl}


\bibitem[Reddy(2011)]%
        {Reddy2011}
\bibfield{author}{\bibinfo{person}{Thomas~B Reddy}.} \bibinfo{year}{2011}\natexlab{}.
\newblock \bibinfo{booktitle}{\emph{{Linden's Handbook of Batteries, 4th Edition}} (\bibinfo{edition}{4th} ed.)}.
\newblock \bibinfo{publisher}{McGraw-Hill Education}, \bibinfo{address}{New York}.
\newblock
\showISBNx{9780071624213}


\bibitem[Rokrok et~al\mbox{.}(2017)]%
        {Rokrok2017}
\bibfield{author}{\bibinfo{person}{Ebrahim Rokrok}, \bibinfo{person}{Miadreza Shafie-khah}, \bibinfo{person}{Pierluigi Siano}, {and} \bibinfo{person}{Jo{\~{a}}o Catal{\~{a}}o}.} \bibinfo{year}{2017}\natexlab{}.
\newblock \showarticletitle{{A Decentralized Multi-Agent-Based Approach for Low Voltage Microgrid Restoration}}.
\newblock \bibinfo{journal}{\emph{Energies}} \bibinfo{volume}{10}, \bibinfo{number}{10} (\bibinfo{date}{sep} \bibinfo{year}{2017}), \bibinfo{pages}{1491}.
\newblock
\showISSN{1996-1073}
\urldef\tempurl%
\url{https://doi.org/10.3390/en10101491}
\showDOI{\tempurl}


\bibitem[Sadeque et~al\mbox{.}(2021)]%
        {Sadeque2021}
\bibfield{author}{\bibinfo{person}{Fahmid Sadeque}, \bibinfo{person}{Dushyant Sharma}, {and} \bibinfo{person}{Behrooz Mirafzal}.} \bibinfo{year}{2021}\natexlab{}.
\newblock \showarticletitle{{Multiple Grid-Forming Inverters in Black-Start: The Challenges}}. In \bibinfo{booktitle}{\emph{2021 IEEE 22nd Workshop on Control and Modelling of Power Electronics (COMPEL)}}. \bibinfo{publisher}{IEEE}, \bibinfo{address}{Cartagena, Colombia}, \bibinfo{pages}{1--6}.
\newblock
\showISBNx{978-1-6654-3635-9}
\urldef\tempurl%
\url{https://doi.org/10.1109/COMPEL52922.2021.9645936}
\showDOI{\tempurl}


\bibitem[Shi et~al\mbox{.}(2015)]%
        {Shi2015}
\bibfield{author}{\bibinfo{person}{Guodong Shi}, \bibinfo{person}{Weiguo Xia}, {and} \bibinfo{person}{Karl~Henrik Johansson}.} \bibinfo{year}{2015}\natexlab{}.
\newblock \showarticletitle{{Convergence of max–min consensus algorithms}}.
\newblock \bibinfo{journal}{\emph{Automatica}}  \bibinfo{volume}{62} (\bibinfo{date}{dec} \bibinfo{year}{2015}), \bibinfo{pages}{11--17}.
\newblock
\showISSN{00051098}
\urldef\tempurl%
\url{https://doi.org/10.1016/j.automatica.2015.09.012}
\showDOI{\tempurl}


\bibitem[Shi et~al\mbox{.}(2019)]%
        {Shi2019}
\bibfield{author}{\bibinfo{person}{Zhichao Shi}, \bibinfo{person}{Hao Liang}, \bibinfo{person}{Shengjun Huang}, {and} \bibinfo{person}{Venkata Dinavahi}.} \bibinfo{year}{2019}\natexlab{}.
\newblock \showarticletitle{{Distributionally robust chance-constrained energy management for islanded microgrids}}.
\newblock \bibinfo{journal}{\emph{IEEE Transactions on Smart Grid}} \bibinfo{volume}{10}, \bibinfo{number}{2} (\bibinfo{year}{2019}), \bibinfo{pages}{2234--2244}.
\newblock
\showISSN{19493053}
\urldef\tempurl%
\url{https://doi.org/10.1109/TSG.2018.2792322}
\showDOI{\tempurl}


\bibitem[Smith and Ton(2013)]%
        {smith2013}
\bibfield{author}{\bibinfo{person}{Merrill Smith} {and} \bibinfo{person}{Dan Ton}.} \bibinfo{year}{2013}\natexlab{}.
\newblock \showarticletitle{Key Connections: The U.S. Department of Energy's Microgrid Initiative}.
\newblock \bibinfo{journal}{\emph{IEEE Power and Energy Magazine}} \bibinfo{volume}{11}, \bibinfo{number}{4} (\bibinfo{year}{2013}), \bibinfo{pages}{22--27}.
\newblock
\urldef\tempurl%
\url{https://doi.org/10.1109/MPE.2013.2258276}
\showDOI{\tempurl}


\bibitem[Stark et~al\mbox{.}(2021)]%
        {stark2021restoration}
\bibfield{author}{\bibinfo{person}{Sanja Stark}, \bibinfo{person}{Anna Volkova}, \bibinfo{person}{Sebastian Lehnhoff}, {and} \bibinfo{person}{Hermann de Meer}.} \bibinfo{year}{2021}\natexlab{}.
\newblock \showarticletitle{Why Your Power System Restoration Does Not Work and What the ICT System Can Do About It}. In \bibinfo{booktitle}{\emph{Proceedings of the Twelfth ACM International Conference on Future Energy Systems}} (Virtual Event, Italy) \emph{(\bibinfo{series}{e-Energy '21})}. \bibinfo{publisher}{Association for Computing Machinery}, \bibinfo{address}{New York, NY, USA}, \bibinfo{pages}{269--273}.
\newblock
\showISBNx{9781450383332}
\urldef\tempurl%
\url{https://doi.org/10.1145/3447555.3465415}
\showDOI{\tempurl}


\bibitem[Vandoorn et~al\mbox{.}(2013)]%
        {Vandoorn2013}
\bibfield{author}{\bibinfo{person}{T.~L. Vandoorn}, \bibinfo{person}{J.~D.M. {De Kooning}}, \bibinfo{person}{B. Meersman}, {and} \bibinfo{person}{L. Vandevelde}.} \bibinfo{year}{2013}\natexlab{}.
\newblock \showarticletitle{{Review of primary control strategies for islanded microgrids with power-electronic interfaces}}.
\newblock \bibinfo{journal}{\emph{Renewable and Sustainable Energy Reviews}}  \bibinfo{volume}{19} (\bibinfo{year}{2013}), \bibinfo{pages}{613--628}.
\newblock
\showISSN{13640321}
\urldef\tempurl%
\url{https://doi.org/10.1016/j.rser.2012.11.062}
\showDOI{\tempurl}


\bibitem[Wang et~al\mbox{.}(2016)]%
        {Wang2016a}
\bibfield{author}{\bibinfo{person}{Yezhou Wang}, \bibinfo{person}{Chen Chen}, \bibinfo{person}{Jianhui Wang}, {and} \bibinfo{person}{Ross Baldick}.} \bibinfo{year}{2016}\natexlab{}.
\newblock \showarticletitle{{Research on Resilience of Power Systems under Natural Disasters - A Review}}.
\newblock \bibinfo{journal}{\emph{IEEE Transactions on Power Systems}} \bibinfo{volume}{31}, \bibinfo{number}{2} (\bibinfo{year}{2016}), \bibinfo{pages}{1604--1613}.
\newblock
\showISSN{08858950}
\urldef\tempurl%
\url{https://doi.org/10.1109/TPWRS.2015.2429656}
\showDOI{\tempurl}


\bibitem[Zhang et~al\mbox{.}(2010)]%
        {Zhang2010}
\bibfield{author}{\bibinfo{person}{Lidong Zhang}, \bibinfo{person}{Lennart Harnefors}, {and} \bibinfo{person}{Hans-peter Nee}.} \bibinfo{year}{2010}\natexlab{}.
\newblock \showarticletitle{{Power-Synchronization Control of Grid-Connected Voltage-Source Converters}}.
\newblock \bibinfo{journal}{\emph{IEEE Transactions on Power Systems}} \bibinfo{volume}{25}, \bibinfo{number}{2} (\bibinfo{date}{5} \bibinfo{year}{2010}), \bibinfo{pages}{809--820}.
\newblock
\showISSN{0885-8950}
\urldef\tempurl%
\url{https://doi.org/10.1109/TPWRS.2009.2032231}
\showDOI{\tempurl}


\bibitem[Zhang et~al\mbox{.}(2016)]%
        {Zhang2016}
\bibfield{author}{\bibinfo{person}{Sai Zhang}, \bibinfo{person}{Cihan Tepedelenlioglu}, \bibinfo{person}{Mahesh~K. Banavar}, {and} \bibinfo{person}{Andreas Spanias}.} \bibinfo{year}{2016}\natexlab{}.
\newblock \showarticletitle{{Max Consensus in Sensor Networks: Non-Linear Bounded Transmission and Additive Noise}}.
\newblock \bibinfo{journal}{\emph{IEEE Sensors Journal}} \bibinfo{volume}{16}, \bibinfo{number}{24} (\bibinfo{date}{dec} \bibinfo{year}{2016}), \bibinfo{pages}{9089--9098}.
\newblock
\showISSN{1530-437X}
\urldef\tempurl%
\url{https://doi.org/10.1109/JSEN.2016.2612642}
\showDOI{\tempurl}


\bibitem[Zhou et~al\mbox{.}(2023)]%
        {Zhou2023}
\bibfield{author}{\bibinfo{person}{Xu Zhou}, \bibinfo{person}{Zhongjing Ma}, \bibinfo{person}{Suli Zou}, \bibinfo{person}{Jinhui Zhang}, {and} \bibinfo{person}{Yonglin Guo}.} \bibinfo{year}{2023}\natexlab{}.
\newblock \showarticletitle{{Distributed Energy Management of Double-Side Multienergy Systems via Sub-Gradient Averaging Consensus}}.
\newblock \bibinfo{journal}{\emph{IEEE Transactions on Smart Grid}} \bibinfo{volume}{14}, \bibinfo{number}{2} (\bibinfo{year}{2023}), \bibinfo{pages}{979--995}.
\newblock
\showISSN{19493061}
\urldef\tempurl%
\url{https://doi.org/10.1109/TSG.2022.3201814}
\showDOI{\tempurl}


\bibitem[Zidan and El-Saadany(2012)]%
        {zidan6205352mas}
\bibfield{author}{\bibinfo{person}{Aboelsood Zidan} {and} \bibinfo{person}{Ehab~F. El-Saadany}.} \bibinfo{year}{2012}\natexlab{}.
\newblock \showarticletitle{A Cooperative Multiagent Framework for Self-Healing Mechanisms in Distribution Systems}.
\newblock \bibinfo{journal}{\emph{IEEE Transactions on Smart Grid}} \bibinfo{volume}{3}, \bibinfo{number}{3} (\bibinfo{year}{2012}), \bibinfo{pages}{1525--1539}.
\newblock
\urldef\tempurl%
\url{https://doi.org/10.1109/TSG.2012.2198247}
\showDOI{\tempurl}


\end{thebibliography}
